\documentclass[10pt,sigplan,screen,final]{acmart}

\setcopyright{acmlicensed}
\copyrightyear{2022}

\acmYear{2022}

\acmConference[CPP '22]{Proceedings of the 11th ACM SIGPLAN International Conference on Certified Programs and Proofs}{January 17--18, 2022}{Philadelphia, PA, USA}
\acmBooktitle{Proceedings of the 11th ACM SIGPLAN International Conference on Certified Programs and Proofs (CPP '22), January 17--18, 2022, Philadelphia, PA, USA}
\acmPrice{15.00}
\acmDOI{10.1145/3497775.3503671}
\acmISBN{978-1-4503-9182-5/22/01}
\acmSubmissionID{}

\usepackage[T1]{fontenc}
\usepackage[utf8]{inputenc}
\usepackage{utf8-symbols}

\usepackage{microtype}
\usepackage{nccmath}
\usepackage{textcomp}
\usepackage{mathtools}
\usepackage{xparse}
\usepackage{bbold}
\usepackage{stmaryrd}
\usepackage{wrapfig}
\usepackage{multirow}
\usepackage{fancyvrb}
\usepackage{xcolor}
\usepackage{float}
\usepackage{graphicx}
\usepackage{amsmath}
\usepackage{amsthm}
\usepackage{dirtytalk}
\usepackage{epigraph}
\usepackage{tikz}
\usepackage{quiver}
\usepackage{subcaption}
\usepackage{relsize}
\allowdisplaybreaks

\definecolor{darkgreen}{HTML}{014801}
\definecolor{darkblue}{HTML}{191970}

\usepackage[all,british]{foreign}

\usepackage{mathpartir}
\usepackage[nameinlink,capitalize,noabbrev]{cleveref}

\usepackage{thmtools}

\newtheorem{remark}{Remark}

\newenvironment{linkproof}[2][]{%
    \begin{proof}[#1]
        
    }{%
    \end{proof}
    
}
{%
        \hfill\href{\linkagdadefinition}{\(\diamond\)}
        \end{definition}
}
{%
        \hfill\href{\linkagdatheorem}{\(\diamond\)}
        \end{theorem}
}
{%
        \hfill\href{\linkagdalemma}{\(\diamond\)}
        \end{lemma}
}
{%
        \hfill\href{\linkagdacorollary}{\(\diamond\)}
        \end{corollary}
}
\numberwithin{equation}{section}

\usetikzlibrary{decorations.markings}
\usetikzlibrary{decorations.pathreplacing}
\usetikzlibrary{matrix}
\usetikzlibrary{arrows}
\usetikzlibrary{chains}
\usetikzlibrary{positioning}
\usetikzlibrary{scopes}
\usetikzlibrary{decorations.pathmorphing}

\usepackage[textsize=footnotesize, obeyFinal]{todonotes}

\newcommand{\Edge}{\ensuremath{\mathsf{E}}}

\newcommand{\Graph}{\ensuremath{\mathsf{Graph}}}

\newcommand{\HomWalk}[3]{\ensuremath{#2\sim_{#1}#3}}

\newcommand{\isSet}[1]{\ensuremath{\mathsf{isSet}(#1)}}

\newcommand{\Node}{\ensuremath{\mathsf{N}}}
\newcommand{\N}{\mathbb{N}}

\newcommand{\UU}{\mathcal{U}}

\graphicspath{{./../reports/ipe-images/}}

\hypersetup{
  unicode=true,
  pdfkeywords={graph maps, walk normal forms, HoTT, Agda},
  pdftitle={On Homotopy of Walks and Spherical Maps in Homotopy Type Theory},
  pdfauthor={J. Prieto-Cubides},
  colorlinks=true,
  linkcolor=darkgreen,
  citecolor=darkgreen,
  urlcolor=darkgreen,
  breaklinks=true
}

\usepackage[nameinlink,capitalize,noabbrev]{cleveref}
\Crefname{theorem}{Theorem}{Theorems}
\Crefname{lem}{Lemma}{Lemmas}
\Crefname{lem}{Corollary}{Corollaries}
\Crefname{listthm}{Theorem}{Theorems}
\Crefname{listlem}{Lemma}{Lemmas}
\crefname{axiom}{Axiom}{Axioms}
\Crefname{axiom}{Axiom}{Axioms}
\Crefname{fun}{Function}{Functions}
\Crefname{eq}{}{}
\Crefname{equiv}{Equivalence}{Equivalences}
\Crefname{equality}{Equality}{Equalities}
\Crefname{cond}{Condition}{Conditions}
\Crefname{calc}{Calculation}{Calculations}
\Crefname{type}{Type}{Types}
\creflabelformat{eq}{#2(#1)#3}
\creflabelformat{type}{#2(#1)#3}
\creflabelformat{equality}{#2(#1)#3}
\creflabelformat{calc}{#2(#1)#3}
\creflabelformat{equiv}{#2(#1)#3}
\Crefname{fig}{Figure}{Figures}
\Crefname{proof}{Proof}{Proofs}

\usepackage{datetime2} 

\usepackage{caption}

\usepackage{flushend}

\begin{document}

\title{On Homotopy of Walks and Spherical Maps in Homotopy Type Theory}
\author{Jonathan Prieto-Cubides}
\orcid{0000-0002-8449-3812}
\affiliation{%
  \department{Department of Informatics}%
  \institution{University of Bergen}%
    \streetaddress{Postboks 7803, Thormøhlens Gate 55}%
      \city{Bergen}%
        \postcode{5020}%
      \country{Norway}%
}
\email{jonathan.cubides@uib.no}%

\begin{abstract}
We work with combinatorial maps
to represent graph embeddings into surfaces up to isotopy. The surface
in which the graph is embedded is left implicit in this approach. The
constructions herein are proof-relevant and stated with a
subset of the language of homotopy type theory.

This article presents a refinement of one characterisation of
embeddings in the sphere, called \emph{spherical maps}, of connected
and directed multigraphs with discrete node sets. A combinatorial
notion of homotopy for walks and the normal form of walks under a
reduction relation is introduced. The first characterisation of
spherical maps states that a graph can be embedded in the sphere if
any pair of walks with the same endpoints are merely walk-homotopic.
The refinement of this definition filters out any walk with inner
cycles. As we prove in one of the lemmas, if a spherical map is
given for a graph with a discrete node set, then any walk in the graph
is merely walk-homotopic to a normal form.

The proof assistant Agda contributed to formalising the results
recorded in this article. 
\end{abstract}

%



\maketitle

\hypertarget{introduction}{%
\section{Introduction}\label{introduction}}

This paper investigates the notion of homotopy of walks to study an
equivalence between two definitions of embeddings in the sphere of
connected and locally finite directed multigraphs. The constructions
are proof-relevant and constructive, powered by homotopy type theory
(HoTT) as the chosen mathematical foundation
\citep{hottbook, Escardo2019}.

The topological graph theory approach inspires our definition of a
combinatorial notion of embedding/map in the sphere for graphs
\citep{planarityHoTT}, referred to as \textit{spherical maps} in this
paper, see \Cref{def:spherical-map}. A graph map can be described by
the graph itself and the circular ordering of the edges incident to
each vertex \citep[§3]{gross}. Using this description, a graph is
understood to be embedded in the sphere if the walks with the same
endpoints are \emph{walk-homotopic}, similar to the topological
concept of a connected closed and simply connected space. We propose a
more pragmatic characterisation of spherical maps, using the fact that
cycles/loops in the graph are walk-homotopic to a point in the sphere.
To prove a map is spherical for a graph with a discrete node set, it
is unnecessary to consider the infinite collection of walks. The set
of walks without inner loops suffices, as we proved in
\Cref{lem:two-spherical-map-definition-are-equivalent}.

To demonstrate our main results, we introduce a reduction relation and
the notion of quasi-simple walks in
\Cref{def:loop-reduction-relation,def:simple-walk}, respectively.
Using this reduction relation, as stated in \Cref{thm:normalisation},
it is possible to define a normal form for walks and prove that every
walk always has a normal form under certain conditions. Additionally,
suppose a spherical map is given for a graph with a discrete node set.
In this case, we provide a normalisation theorem to state that any
walk is merely walk-homotopic to a normal form, see the details in
\Cref{thm:hom-normalisation}.

\paragraph{Outline}

The terminology and notation used throughout the paper is presented in
\Cref{sec:background}. Readers familiar with HoTT may want to skip
this section. The type of graphs discussed in this paper is defined in
\Cref{sec:graph-background}. In \Cref{sec:walks}, we define the type
of walks and the type of quasi-simple walks to introduce the normal
form of a walk in \Cref{sec:loop-reduction-relation}. In
\Cref{sec:homotopy-normalisation}, a normalisation theorem for walks
is given. Related work is reviewed in \Cref{sec:related-work}, and
finally, conclusions are drawn and future work outlined in
\Cref{sec:conclusions}.

\paragraph{Computer Formalisation}

One advantage of using dependent type theories, as in this paper, is
checking the correctness of the mathematical constructions using
computer assistance. A proof assistant is a system with support to
write such programs/proofs. The results in this document were
formalised in the proof assistant Agda v(2.6.2), in a fully
self-contained development, which does not depend on any library. The
digital version of this document contains links to the Agda terms for
some definitions, lemmas, and proofs. For example, we have made
clickable the QED symbol
(\href{https://jonaprieto.github.io/synthetic-graph-theory/CPP2022-paper.html}{$\square$})
at the end of a proof.

In the implementation, the formalisation is type-checked using the
flag \texttt{without-K} for compatibility with HoTT \citep{COCKX2016}.
Also, the flag \texttt{exact-split} was used to ensure that all
clauses in a definition are definitional equalities. In our Agda
library, to support this development, we required only a postulate for
function extensionality and the corresponding postulates related to
propositional truncation.

\section{Mathematical Foundation}
\label{sec:background}

Homotopy type theory (HoTT) is an intensional Martin-Löf type theory
(MLTT) \citep{hottbook, Awodey2012} containing Voevodsky's
\emph{Univalence
axiom} \citep{voevodsky2014equivalence} and some higher inductive
types, such as propositional truncation.

Revealed thanks to the formalisation, only a subset of HoTT is
required for the results of this work. Precisely, we only need MLTT
with universes, function extensionality and propositional truncation.
However, since this work is part of a more ambitious project in which
the whole theory is used, let us say that HoTT is our mathematical
foundation for studying graph theory. This approach gives us, for
example, the correct encoding of the equality between graphs, in the
sense of the identity type, coinciding with the notion of graph
isomorphism.

In HoTT, there is a natural correspondence between homotopy theory and
the higher structure of the identity type of intensional MLTT. A space
is a type where points are terms of their corresponding type, and
paths from \(a\) to \(b\) are of the identity type between \(a\) and
\(b\). By such a correspondence, one can, for example, study synthetic
homotopy theory, as presented in the HoTT Book \citep[§8]{hottbook}.

An informal type theoretical notation derived from the HoTT book
\citep{hottbook} and the formal system Agda \citep{norell2007} is used
throughout the paper. Definitions are introduced by (\(:≡\)) while
judgmental equalities use (\(≡\)). The identity type is denoted by
(\(=\)). The universe is denoted by \(\UU\). The notation \(A : \UU\)
indicates that \(A\) is a type. To state that \(a\) is of type \(A\)
we write \(a : A\). The universe \(\UU\) is closed under the following
type formers. The coproduct of two types, \(A\) and \(B\), is denoted
by \(A + B\). The corresponding data constructors are the functions
\(\mathsf{inl} : A \to A + B\) and \(\mathsf{inr}: B \to A+B\). The
dependent sum type (Σ-type) is denoted by \(Σ_{x:A}B(x)\). The
dependent product type (Π-type) is denoted by \(Π_{x:A}B(x)\). The
empty type and unit type are denoted by \(\mathbb{0}\) and
\(\mathbb{1}\), respectively. The type \(x \neq y\) denotes the
function type \((x = y) \to \mathbb{0}\). Natural numbers are of type
\(\N\). \(0 : \N\). The successor of \(n : \N\) is denoted by \(S(n)\)
or \(n+1\). Given \(n : \N\), the type with \(n\) elements is denoted
by \(⟦n⟧\) and is defined inductively by setting
\(⟦0⟧ :\equiv \mathbb{0}\), \(⟦1⟧ :\equiv \mathbb{1}\) and
\(⟦n+1⟧ :\equiv ⟦n⟧ + \mathbb{1}\). To define some inductive types, we
adopt a similar notation as in Agda, including the keyword
\(\mathsf{data}\) and the curly braces for implicit arguments,
e.g.~\(\{a : A\}\) denotes \(a\) is of type \(A\), and it is an
implicit variable. The type may be omitted in the former notation, as
they can usually be inferred from the context.

We follow the HoTT Book, with slight changes in notation, for
definitions such as embeddings, equivalence of types denoted by
\((\simeq)\), propositional truncation of type \(A\) denoted by
\(\|A\|\), and \(n\)-types, e.g.~contractible types, propositions, and
sets, with their corresponding predicate, \(\mathsf{isContr}\),
\(\mathsf{isProp}\), and \(\mathsf{isSet}\).

\begin{theorem}[Hedberg's theorem]\label{lem:hedberg}
A type $A$ with decidable equality, i.e.  $x = y$ or $x \neq y$
for all $x,y : A$, forms a set, and it is below referred to as \emph{discrete} set.
\end{theorem}

It remains to define two fundamental notions towards studying the
combinatorics of graphs, namely the type of finite sets and cyclic
sets.

\begin{definition}\label{def:finite-type}
Given $X : \UU$, let $\mathsf{isFinite}(X) : \UU$ be given by
\begin{equation}\label[type]{def:finite}
\mathsf{isFinite}(X) :≡ \sum_{(n~:~\N)} \left\| X \simeq \llbracket n \rrbracket \right  \|.
\end{equation}
\end{definition}

The finiteness of a type \(X\) is the existence of a bijection between
\(X\) and the type \(⟦n⟧\) for some \(n:\mathbb{N}\). One can prove
that \Cref{def:finite} is a proposition. A type \(X\) is called
\emph{finite} if \(\mathsf{isFinite}(X)\) holds. The corresponding
natural number \(n\) is referred as the cardinal number of \(X\). Any
property on \(⟦n⟧\), for example, \say{being a set} and
\say{being discrete}, can be transported to any finite type.

\begin{lemma}\label{lem:finiteness-closure-property} Finite sets are
closed under (co) products, type equivalences, $\Sigma$-types and
$\Pi$-types. 
\end{lemma}

For example, if \(A\) is a finite set and \(B : A \to \UU\) is a type
family such that for each \(a:A\) the type \(B(a)\) is a finite set,
one can conclude that the type \(\Pi_{x:A}\,B(x)\) is a finite set.
The formal proof of \Cref{lem:finiteness-closure-property} and other
\href{http://hott.github.io/HoTT/coqdoc-html/HoTT.Spaces.Finite.Finite.html}{related lemmas}
can be found in the Coq-HoTT library \citep{HoTTCoq}. For example, one
of such lemmas, used to demonstrate \Cref{lem:lemma0}, states that the
cardinality of \(X\) is less than or equal to the cardinality of \(Y\)
if there exists an embedding from \(X\) to \(Y\).

As the very first examples of finite sets, we have the empty type,
unit type, decidable propositions and the family of types \(⟦n⟧\) for
every \(n:\mathbb{N}\). To prove the finiteness of other types, as in
\Cref{thm:finite-simple-walks}, we use
\Cref{lem:decidable-implies-finite-path}, a direct consequence of
Hedberg's theorem and finiteness of the empty and unit type.

\begin{lemma}\label{lem:decidable-implies-finite-path} If $A$ is
discrete, then the identity type $x = y$ is a finite set for all
$x,y:A$. 
\end{lemma}

We now present a definition of cyclic types, used later to define the
combinatorial characterisation of graphs embedded in a surface in
\Cref{def:graph-map}. Being cyclic for a type is a structure, not a
property, given by preserving the structure of cyclic subgroups of
permutations on \(⟦n⟧\). To endow a type with such a cyclic structure,
let \(\mathsf{pred}\) be the predecessor function of type
\(⟦n⟧ → ⟦n⟧\), defined as the mapping, \(0↦ (n-1)\) and
\((m+1) \mapsto m\) for \(m < n\).

\begin{definition}\label{def:cyclic-type}
Given $A : \UU$, we define the type of \emph{cyclic structures}
on $A$, $\mathsf{Cyclic}(A)$, as follows.

\begin{equation*}
\mathsf{Cyclic}(A) :≡ \sum_{(\varphi~:~A → A)}\sum_{(n~:~\N)} \| ∑_{(e~:~A\,≃\,⟦n⟧)} 
(e ∘ \varphi = \mathsf{pred} ∘ e) \|.
\end{equation*}

A cyclic structure is denoted by a tuple $⟨\varphi, n⟩$ where
$(\varphi, n, p)$ is of type $\mathsf{Cyclic}(A)$. One may omit $n$
for brevity if no confusion arises. A type $A$ with a cyclic structure
$⟨\varphi, n⟩$ is referred as an $n$-cyclic type or simply as a
\emph{cyclic set} with $n$ elements.

\end{definition}

\section{The Type of Graphs}\label{sec:graph-background}

A \emph{graph} is a term of the type in \Cref{def:graph}. The
corresponding data is a set of \emph{nodes} and a set for each pair of
nodes called \emph{edges}.

\begin{definition}\label{def:graph} A directed multigraph is of the following
type.

    \begin{equation*}
        \Graph :≡ \hspace{-2mm}\sum_{(\Node~:~\UU)}\sum_{(\Edge~:~\Node → \Node →
        \UU)}\hspace{-2mm}\isSet{\Node} × \prod_{(x,y~:~\Node)} \isSet{\Edge(x,y)}.
    \end{equation*}
\end{definition}

Given a graph \(G\), the set of nodes is denoted by \(\Node_{G}\).
Given two nodes \(x\) and \(y\), the edges between them form a set
denoted by \(\Edge_{G}(x,y)\). If \(e\) is an edge from \(x\) to
\(y\), we denote by \(\mathsf{source}(e)\) the node \(x\) and by
\(\mathsf{target}(e)\) the node \(y\). A \emph{finite graph} is a
graph where the node set is a finite set as well as every family of
sets \(\Edge_{G}(x,y)\). One can prove that the type of graphs in
\Cref{def:graph} forms a homotopy groupoid and is also a univalent
category \citep{hottbook}. The proof of these facts and
\href{https://jonaprieto.github.io/synthetic-graph-theory/lib.graph-definitions.Graph.EquivalencePrinciple.html}{related lemmas}
will be omitted as it is not essential for our work here. The
interested reader can check the formalisation in Agda for the
respective proofs \citep{agdaformalisation}. In the upcoming sections,
unless stated otherwise, we will denote \(G\) to be a graph, and
\(x,y\), and \(z\) to be variables for nodes in \(G\).

\section{Walks in a Graph}\label{sec:walks}

The notion of a walk plays an essential role in graph theory. Many of
the algorithms using graph data structures are based on this object.
One may be interested in finding the \say{distance between two nodes}
in a graph, the shortest walk, and several other variation problems
related to walking in the graph.

\begin{definition}\label{def:walk} A \emph{walk} in $G$ from $x$ to
$y$ is a sequence of connected edges that we construct using the
following inductive data type:

\begin{equation*}
\begin{aligned}
\mathsf{\textbf{data}} & \; \mathsf{W}\,~:~\, \Node_{G} \to \Node_{G} \to \UU\\
& \langle\_\rangle \,:\, (x~:~\Node_{G}) \to \mathsf{W}_{G}(x,x)\\
& (\_\hspace{-1mm}\odot\hspace{-1mm}\_) \,:\, \Pi\,\{x\,y\,z~:~\Node_{G}\}\,.\,(e~:~\Edge_{G}(x,y))\\
& \hspace{8mm} \to (w~:~\mathsf{W}_{G}(y,z))\\
& \hspace{8mm} \to \mathsf{W}_{G}(x,z)
\end{aligned}
\end{equation*}

Let $w$ be a walk from $x$ to $y$, i.e.  of type $\mathsf{W}_{G}(x,y)$.
We will denote by $x$ the \emph{head} of $w$ and by $y$ the \emph{end}
of $w$. If $w$ is $\langle x\rangle$ then we refer to $w$ as
\emph{trivial} or \emph{one-point} walk. If $w$ is of the form $(e ⊙
\langle x \rangle)$, then $w$ is the \emph{one-edge} walk $e$.
Nontrivial walks are of the form $(e⊙w)$ and a \emph{loop} is a walk
with the same head and end.
\end{definition}

\hypertarget{structural-induction-for-walks}{%
\subsection{Structural Induction for
Walks}\label{structural-induction-for-walks}}

By \emph{structural induction} or \emph{pattern matching} on a walk,
we will refer to the elimination principle of the inductive type in
\Cref{def:walk}. An induction principle allows us to define outgoing
functions from a type to a type family. For instance, if we want to
use the induction principle to inhabit a predicate on the type of
walks, \(P:\Pi\{x\,y:\Node_{G}\}.\mathsf{W}_{G}(x,y) \to \UU\), one
can inhabit \Cref{eq:structural-induction}. Given a walk
\(w:\mathsf{W}_{G}(x,y)\), to construct a term of type \(P(w)\), the
base case must first be constructed, i.e.~give a term of type
\(P(\langle x \rangle)\), for every \(x~:~\Node_{G}\). Subsequently,
we must prove the case for composite walks, i.e.~\(P(e ⊙ w)\). To show
this, \(P(w)\) is assumed for any walk \(w\), and we construct a term
of type \(P(e ⊙ w)\) from this assumption. Thus, one gets \(P(w)\) for
any walk \(w\). Another induction principle for walks is stated in
\Cref{thm:walk-induction-by-length}.

\begin{equation}\label[type]{eq:structural-induction}
 \begin{split}
    & \hspace{3mm}{\prod_{(x~:~\Node_{G})}\,P(\langle x \rangle)}\, \\
    & {\times  \prod_{(x, y ,z~:~\Node_{G})} \prod_{(e~:~\Edge_{G}(x,y))}
     \prod_{(w~:~\mathsf{W}_{G}(y,z))} P(w) \to P(e ⊙ w)} \\
    &{\to \prod_{(x,y~:~\Node_{G})} \prod_{(w~:~\mathsf{W}_{G}(x,y))} P(w)}.
 \end{split}
\end{equation}

The \emph{composition}, also called concatenation, of walks is an
associative binary operation on walks defined by structural induction
on its left argument. Given walks \(p : \mathsf{W}_{G}(x,y)\) and
\(q : \mathsf{W}_{G}(y,z)\), we refer to their composition as the
\emph{composite} denoted by \(p \cdot q\). The node \(y\) is called
the \emph{joint} of the composition. The \emph{length} of the walk
\(w\) is denoted by \(\mathsf{length}(w)\) and represents the number
of edges used to construct \(w\). A trivial walk has length zero,
whilst a walk \((e ⊙ w)\) has one more length than \(w\). We display a
point to represent trivial walks and with a normal arrow to represent
walks of positive length, as illustrated in \Cref{fig:simple-walk}.

\begin{lemma}\label{lem:walk-is-set} The type of walks forms a set.
\end{lemma}

\begin{linkproof}[]{https://jonaprieto.github.io/synthetic-graph-theory/CPP2022-paper.html\#1201}
One can show that the type $\mathsf{W}(x,y)$ is
\href{https://jonaprieto.github.io/synthetic-graph-theory/lib.graph-walks.Walk.SigmaWalks.html}{equivalent}
 to $\Sigma_{n :\mathbb{N}}\,\hat{W}(n,x,y)$ with $\hat{W}$
defined as follows.
\begin{subequations}\label{def:walk-2}
    \begin{align}
        &\hat{W}~:~\mathbb{N} \to \Node_{G} \to \Node_{G} \to \UU\\
        &\hat{W}(0,x,y) :\equiv (x = y),\label{eq:w2-zero}\\
        &\hat{W}(S(n),x,y) :\equiv \sum_{(k~:~\Node_{G})}\,\Edge_{G}(x,k) \times \hat{W}(n,k,y) \label{eq:w2-sn}.
    \end{align}
\end{subequations}

It suffices to show that the type $\hat{W}(n,x,y)$ forms a set for
$n:\mathbb{N}$ which will be proven by induction on $n$. If $n=0$, one
obtains the proposition $x = y$ which is a set.
Consequently, we must now show that the type in \Cref{eq:w2-sn} is a
set. By the graph definition, the base type $\Node_{G}$ and
$\Edge_{G}$ are both sets. Thus, one only requires that $\hat{W}(n,
k,y)$ forms a set, which is precisely the induction hypothesis.
\end{linkproof}

Although it is not included in the formalisation of this work, one can
show that the type of walks forms a category. If \(\mathsf{Graph}\) is
the category of graphs using \Cref{def:graph} and \(\mathcal{C}\) is
the category of small categories. There is a functor
\(R~:~\mathsf{Graph} \to \mathcal{C}\) mapping every graph \(G\) to
its \emph{free} pre-category. The object set of \(R(G)\) is
\(\Node_{G}\), and the morphisms correspond to the collection of all
possible walks in \(G\). By \cref{lem:walk-is-set}, it follows that
\(R(G)\) is a small category. Let \(L\) be the forgetful functor from
\(\mathcal{C}\) to \(\mathsf{Graph}\). Then, \(L\) is the left adjoint
of \(R\). The \emph{graph of
walks} \(W(G)\) of \(G\) is given by the endofunctor
\(W~:~\mathsf{Graph} \to \mathsf{Graph}\), the monad from the
composite \(L \circ R\).

\subsection{A Well-Founded Order for Walks}\label{sec:well-founded-walks}

Structural induction is a particular case of a more general induction
principle to define recursive programs called \emph{well-founded} or
Noetherian induction. Recall that for the structural induction
principle, one must always guarantee that every argument in a
recursive call in the program is strictly smaller than its arguments.
However, there is no reason to believe this will always be the case.

In constructive mathematics, a binary relation \(R\) on a set \(A\) is
\emph{well-founded} if every element of \(A\) is \emph{accessible}. An
element \(a:A\) is accessible by \(R\), if \(b:A\) is accessible for
every \(bRa\) \citep[§10.3]{Nordstrm1988, hottbook}. Then, if \(a\)
has the property that there is no \(b\) such that \(bRa\), then \(a\)
is vacuously accessible. If (\(\leq\)) represents the
\emph{less or equal than} relation on the natural numbers, then the
number zero is vacuously accessible by \(\leq\) on \(\mathbb{N}\).

Let us define a well-founded order for walks in a graph by considering
their lengths, from where the well-founded induction for walks
follows, see \Cref{thm:walk-induction-by-length}.

\begin{definition}\label{def:walk-order} Given $p,q :
    \mathsf{W}_{G}(x,y)$ for $x,y:\Node_{G}$, the relation
    $(\preccurlyeq)$ states that $p \preccurlyeq q$ when
    $\mathsf{length}(p)\leq \mathsf{length}(q)$. 
\end{definition}

\begin{lemma}\label{lem:well-founded-walk-relation} The relation
($\preccurlyeq$) on $\Sigma_{x,y~:~\Node_{G}} \mathsf{W}_{G}(x,y)$ is well-founded.
\end{lemma}
\begin{linkproof}[]{https://jonaprieto.github.io/synthetic-graph-theory/CPP2022-paper.html\#1278}
It follows from the fact that the poset $(\mathbb{N}, \leq)$ is well-founded.
\end{linkproof}

We refer to the following
\href{https://jonaprieto.github.io/synthetic-graph-theory/CPP2022-paper.html\#1298}{lemma}
as the \emph{well-founded induction principle for walks} induced by
\Cref{def:walk-order}.

\begin{theorem}
    \label{thm:walk-induction-by-length}
    Suppose the following is given,
\begin{enumerate}
\item
a predicate $P$ of type $\Sigma_{x,y~:~\Node_{G}} \mathsf{W}_{G}(x,y) \to \UU$ such that,
\item
given $(a,b,q)$ of type $\Sigma_{x,y~:~\Node_{G}} \mathsf{W}_{G}(x,y)$, if
$P(p)$ for each walk $p~:~\mathsf{W}_{G}(x',y')$ with $x',y': \Node_{G}$ and $p
\preccurlyeq q$, then $P(a,b,q)$.
\end{enumerate}
Then, given any walk $w : \mathsf{W}_{G}(x,y)$ and $x,y : \Node_{G}$, we
have $P(x,y,w)$.
\end{theorem}

\begin{remark} The induction principle stated in
\Cref{thm:walk-induction-by-length} using
\Cref{lem:well-founded-walk-relation} is equivalent to performing
induction on the length of the walk.
\end{remark}

\Cref{thm:normalisation,thm:hom-normalisation} define algorithms for
which many of their recursive calls are on subwalks of the input walk.
A \emph{subwalk} of a walk \(w\) is a contiguous subsequence of edges
in \(w\). Subwalks are not structurally smaller than their
corresponding walk, unless one takes for example the subwalk \(w\) or
\(e\) for the composite walk \((e ⊙ w)\). Excluding the previous case,
to deal with other subwalk cases, we can use the \emph{well-founded}
induction principle given in \Cref{thm:walk-induction-by-length}.

\subsection{Quasi-Simple Walks}\label{sec:quasi-simple}

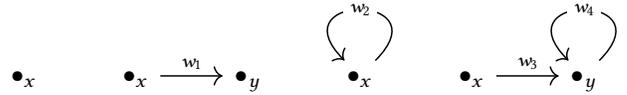
\begin{figure}[!ht]
\centering
\begin{equation*}
    \begin{tikzcd}
        & && \\
        \bullet_{x} &\bullet_{x} &\bullet_{y} &\bullet_{x} & \bullet_{x}  & \bullet_{y}
        \arrow["w_1", from=2-2, to=2-3]
        \arrow["w_2"{description}, from=2-4, to=2-4, loop]
        \arrow["w_3", from=2-5, to=2-6]
        \arrow["w_4"{description},from=2-6, to=2-6, loop]
    \end{tikzcd}
\end{equation*}

\caption{The arrows in the picture can represent edges or walks of a
positive length. In the sense of \Cref{def:simple-walk}, a
quasi-simple walk can only be one of these kinds: i) one-point walk
ii) path iii) loop without inner node repetitions, or iv) composite
walk between a path and a quasi-simple walk of kind iii. The walks
$w_3$ and $w_4$ only share the node $y$.}
\label{fig:simple-walk}
\end{figure}

In this subsection, we characterise walks with shapes as in
\Cref{fig:simple-walk} and refer to such as \emph{quasi-simple} walks
in \Cref{def:simple-walk}.

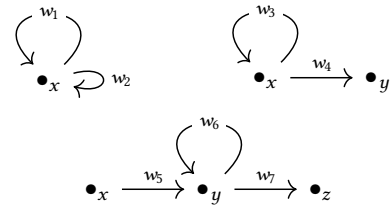
\begin{figure}[!ht]
\centering
\begin{subfigure}[b]{0.33\linewidth}
    \begin{tikzcd}
        \\
        \bullet_{x} 
        \arrow["w_1"{description},from=2-1, to=2-1, loop]
        \arrow["w_2",from=2-1, to=2-1, loop right]
    \end{tikzcd}
\end{subfigure}
\begin{subfigure}[b]{0.33\linewidth}
    \begin{tikzcd}
    &\\
        \bullet_{x}  & \bullet_{y}
        \arrow["w_3"{description},from=2-1, to=2-1, loop]
        \arrow["w_4",from=2-1, to=2-2]
    \end{tikzcd}
\end{subfigure}
\begin{subfigure}[b]{0.33\linewidth}
    \begin{tikzcd}
    & &\\
    \bullet_{x} &\bullet_{y} &\bullet_{z}
        \arrow["w_5",from=2-1, to=2-2]
        \arrow["w_6"{description},from=2-2, to=2-2, loop]
        \arrow["w_7",from=2-2, to=2-3]
    \end{tikzcd}
\end{subfigure}
    
\caption{These are three examples of walks that are not quasi-simple
in the sense of \Cref{def:simple-walk}. The walks $w_1$ and $w_2$ only
share the node $x$, and the same happens with the walks $w_3$ and
$w_4$. The walks $w_5,w_6$ and $w_7$ only share the node $y$.
The walks $w_i$ for $i$ from $1$ to $7$ are nontrivial walks.
}
\label{fig:simple-walk-2}
\end{figure}

The notion of a quasi-simple walk will be used to introduce a
reduction relation on the set of walks to remove their inner loops,
see \Cref{def:loop-reduction-relation}. A related notion to the
quasi-simple walk definition is that of a path \citep{diestel}. The
usual graph-theoretical notion of a \emph{path} is a walk with no
repeated nodes. Here, quasi-simple walks are introduced since paths
are not suitable in our description of graph maps in
\Cref{sec:homotopy-walks-in-sphere}. There, the totality of walks is
considered, which includes closed walks, also called loops. For graph
maps in the sphere, we found out that the type of walks can be
replaced by the type of quasi-simple walks under certain conditions.
Quasi-walks are conveniently defined in a way that permits their end
to appear at most twice in the walk.

To define quasi-simpleness for walks, we introduce a unconventional
relation, denoted by \((x \in w)\), meaning that the node \(x\) is in
the walk \(w\) and it is not the last, see \Cref{def:node-membership}.
\((x \in w)\) is a proposition, and decidable if the walks belong to
graphs with discrete node set. Consequently,
\Cref{lem:being-simple-is-prop} shows that being quasi-simple is also
a decidable proposition on the same kind of graphs. Quasi-simple walks
play a relevant role in this work. They are required to give an
alternative definition of graph maps in the sphere, as stated in
\Cref{def:spherical-map-simple}.

\begin{definition}\label{def:node-membership} Let $x,y,z:\Node_{G}$
    and $w: \mathsf{W}_{G}(x,z)$. The relation $(\in)$ on a walk $w$
    for a node $y$ is defined as the node $y$ that is not $z$ but
    belongs to $w$, i.e. whenever the type $(y \in w)$ is inhabited.

\begin{enumerate}
\item $y \in ⟨z⟩ :\equiv \mathbb{0}$.
\item $y \in (e ⊙ w) :\equiv (y = \mathsf{source}(e)) + (y \in w)$.
\end{enumerate}
\end{definition}

\begin{lemma} If the node set of the graph $G$ is discrete, then the
type $(x ∈ w)$ is decidable proposition for any node $x$ and walk $w$
in $G$.
\end{lemma}

\begin{definition}\label{def:simple-walk} Given $x,y: \Node_{G}$, a walk in $G$ from $x$ to $y$ is \emph{quasi-simple} if $\mathsf{isQuasi}(w)$ holds.
  \begin{equation}\label{eq:simple-walk}
    \mathsf{isQuasi}(w) :≡ \prod_{(z~:~\Node_{G})} \mathsf{isProp}(z \in w).
  \end{equation}
\end{definition}

\begin{lemma}\label{lem:simple-is-prop}
Being quasi-simple is a proposition.
\end{lemma}
\begin{linkproof}[]{https://jonaprieto.github.io/synthetic-graph-theory/CPP2022-paper.html\#1423}
It follows since $\mathsf{isProp}(z \in w)$ is a proposition.
\end{linkproof}

Thus, \Cref{def:simple-walk} presents a quasi-simple walk as a path
where the end could only be present at most twice. Examples of walks
that are not quasi-simple are illustrated in \Cref{fig:simple-walk-2}.

\begin{lemma}\label{lem:e-simple-is-simple} Given $x,y,z : \Node_{G}$,
$e : \Edge_{G}(x,y)$ and a quasi-simple walk $w : \mathsf{W}_{G}(y,z)$, if
$x~\not \in~w$ then the walk $(e ⊙ w)$ is quasi-simple.
\end{lemma}

\begin{linkproof}[]{https://jonaprieto.github.io/synthetic-graph-theory/CPP2022-paper.html\#1439}
Given a node $r$, we must show that $r \in (e ⊙ w)$ is a
proposition. That is equivalent to showing that the type
$(r = x) + (r \in w)$ is a proposition. The coproduct of mutually
exclusive propositions is a proposition. Then, remember that $r = x$
is a given proposition and that the type $(r \in w)$ is also a
proposition since the walk $w$ is quasi-simple by hypothesis. Thus, it
remains to show that there is no term $(p, q)$ where $p: (r = x)$ and
$ q : (r \in w) $. A contradiction arises, since by hypothesis $x~\not
\in~w$ but from $\mathsf{tr}^{λ z \to z \in w}(p)(q) : x \in w$.
\end{linkproof}

\begin{lemma}\label{lem:conservation-simple-walks} Given $x,y,z :
\Node_{G}$, $e:\mathsf{Edge}_{G}(x,y)$, and a walk $w:\mathsf{W}_{G}(y,z)$, if
the walk $(e~⊙~w)$ is a quasi-simple walk then $w$ is also a quasi-simple walk.
\end{lemma}
\begin{linkproof}[]{https://jonaprieto.github.io/synthetic-graph-theory/CPP2022-paper.html\#1481}
Given any node $u~:~\Node_{G}$ and two proofs $p,q~:~u \in w$, we must
show that $p=q$. By definition, $\mathsf{inr}(p)$ and $\mathsf{inr}(q)
$ are proofs that $u \in (e~⊙~w)$. Because $(e~⊙~w)$ is a quasi-simple walk,
the equality $\mathsf{inr}(p) = \mathsf{inr}(q)$ holds. The
constructor $\mathsf{inr}$ is an injective function, and one therefore
obtains $p=q$ as required.\qedhere
\end{linkproof}

\begin{corollary}\label{lem:basic-simple-walks} Trivial and one-edge
walks are quasi-simple walks.
\end{corollary}

\begin{lemma}\label{lem:being-simple-is-prop}
If the node set of the
graph is discrete, then being quasi-simple for a walk is a decidable
proposition.
\end{lemma}

\begin{linkproof}[]{https://jonaprieto.github.io/synthetic-graph-theory/CPP2022-paper.html\#1604}
Let $x,z: \Node_{G}$ and $w : \mathsf{W}_{G}(x,z)$, we want to show
that $\mathsf{isQuasi}(w)$ is decidable. The proof is by induction on
the structure of $w$.
\begin{enumerate}
\item If $w$ is trivial then, by \Cref{lem:basic-simple-walks},
the walk $w$ is quasi-simple. 
\item If $w$ is the composite walk $(e ⊙ w')$ for $e : \Edge_{G}(x,y)$
and $w'~:~\mathsf{W}_{G}(y,z)$, we recursively ask whether the walk
$w'$ is quasi-simple or not. 
\begin{enumerate} 
\item If $w'$ is not quasi-simple, then $w$ is not quasi-simple by
    the contrapositive of \Cref{lem:conservation-simple-walks}. 
\item If $w'$ is quasi-simple, then we ask if $x \in w'$. If so, then $w$ is
not quasi-simple. Otherwise, that would contradict the quasi-simpleness definition,
as the node $x$ would appear twice in $w$. Now, if $x \not \in w'$,
one obtains that $w$ is quasi-simple by \cref{lem:e-simple-is-simple}. \qedhere
\end{enumerate}
\end{enumerate}
\end{linkproof}

\hypertarget{a-finiteness-property}{%
\subsection{A Finiteness Property}\label{a-finiteness-property}}

The goal in this subsection is to prove that the collection of
quasi-simple walks in a finite graph \(G\) forms a finite set, as
stated in \Cref{thm:finite-simple-walks}. To show this, a proof on the
finiteness of an equivalent type to \Cref{def:simple-walk-collection}
is given. To establish such equivalence, see \Cref{lem:lemma1}, we
first need to demonstrate some intermediate results as the following.

\begin{equation}\label[type]{def:simple-walk-collection}
\sum_{(w~:~\mathsf{W}_{G}(x,y))}  \mathsf{isQuasi}(w).
\end{equation}

\begin{lemma}\label{lem:number-of-nodes-in-walk}
    
Given any walk $w : \mathsf{W}_{G}(x,z)$ of length $n$, then

\begin{equation}\label[equiv]{eq:number-of-nodes-in-walk}
   ⟦ n ⟧ \simeq \sum_{(y~:~\Node_{G})} (y ∈ w).
\end{equation}
\end{lemma}

\begin{linkproof}[]{https://jonaprieto.github.io/synthetic-graph-theory/CPP2022-paper.html\#1651}
By induction on the structure of $w$.
\begin{enumerate}
    \item If the walk is trivial, the required equivalence follows
    from the type equivalence between $\mathbb{0}$ and $\Sigma_{z
    :\Node_{G}}\mathbb{0}$.
    \item If the walk is $(e ⊙ w)$ for $e~:~\Edge_{G}(x,y)$ and $w :
    \mathsf{W}_{G}(y,z)$, the equivalence is established by the
    following calculation. Let $n$ be the length of $w$.

    \begin{subequations}
      \begin{align}
        \sum_{(y~:~\Node_{G})} (y \in (e ⊙ w )) &\equiv  \sum_{(y~:~\Node_{G})} (y = x) + (y \in w ) \label[equiv]{eq:number-nodes-1}\\
        &\simeq  \sum_{(y~:~\Node_{G})} (y = x) + \sum_{(y~:~\Node_{G})} (y \in w )  \label[equiv]{eq:number-nodes-2}\\
        &\simeq \mathbb{1} +  \sum_{(y~:~\Node_{G})} (y \in w ) \label[equiv]{eq:number-nodes-3}\\
        &\simeq  \mathbb{1} +  ⟦ n ⟧  \label[equiv]{eq:number-nodes-4} \\
        &\simeq  ⟦ n + 1 ⟧. \label[equiv]{eq:number-nodes-5}
      \end{align}
    \end{subequations}
    \cref{eq:number-nodes-1} is accomplished by
    \Cref{def:node-membership}. $\Sigma$-type distributes coproducts
    as in \Cref{eq:number-nodes-2}. We can simplify in
    \Cref{eq:number-nodes-3} because the type $\Sigma_{y:\Node_{G}} (y
    = x)$ is contractible. Note that the inner path is fixed and it is
    then equivalent to the unit type. \Cref{eq:number-nodes-4} is by
    the induction hypothesis applied to $w$. \Cref{eq:number-nodes-5}
    is accomplished by the definition of $⟦ n ⟧$ using the coproduct
    definition. \qedhere
\end{enumerate}
\end{linkproof}

\begin{lemma}\label{lem:inw-is-finite} Given $x,y,z : \Node_{G}$, and
    $w : \mathsf{W}_{G}(x,y)$ the type $(z ∈ w)$ is a finite set if
    the node set of $G$ is discrete.
\end{lemma}

\begin{linkproof}[]{https://jonaprieto.github.io/synthetic-graph-theory/CPP2022-paper.html\#1680}
By induction on the structure of $w$: in case the walk is trivial, the
type in question is finite as it is equal to the empty type by
definition. In the composite walk case, $z \in (e ⊙ w)$, we must prove
that the type $(z = x) + (z \in w)$ is finite. Note that the former is
finite by \Cref{lem:decidable-implies-finite-path}. By the induction
hypothesis: the type $z \in w$ is finite.  The required conclusion
then follows since finite sets are closed under coproducts. \qedhere
\end{linkproof}

We can now prove that for finite graphs there exists a finiteness
property for the collection of all quasi-simple walks, derived from
the finiteness of the set of quasi-simple walks of a fixed length
\(n\) for \(n:\mathbb{N}\).

\begin{definition}\label{def:finite-simple-walks}
Given $x,y : \Node_{G}$ and $n:\mathbb{N}$, the 
type $\mathsf{qswalk}$ collects all quasi-simple walks of a
fixed length $n$.
\begin{equation*}
\mathsf{qswalk}(n,x,y):\equiv\hspace{-3mm} \sum_{(w~:~\mathsf{W}_{G}(x,y))}
\hspace{-3mm}\mathsf{isQuasi}(w)~\times~(\mathsf{length}(w)=n).
\end{equation*}
\end{definition}

\begin{lemma}\label{lem:equiv-type-fswalk} Given a graph $G$, $n~:~\mathbb{N}$, and
$x, z~:~\Node_{G} $, the following equivalence holds.
\begin{equation}\label[equiv]{eq:equiv-type-fswalk}
\mathsf{qswalk}(S(n), x, z)
\simeq\hspace{-3mm}\sum_{(y~:~\Node_{G})}
\sum_{(e~:~\Edge_{G}(x,y))}\sum_{(w~:~\mathsf{qswalk}(n, y, z))}\hspace{-3mm} (x \not \in w).
\end{equation}
\end{lemma}

\begin{linkproof}[]{https://jonaprieto.github.io/synthetic-graph-theory/CPP2022-paper.html\#1746}
The back-and-forth functions are extensions of the functions derived
from \Cref{lem:e-simple-is-simple,lem:conservation-simple-walks}. \qedhere 
\end{linkproof}

\begin{lemma}\label{lem:lemma2} Given a finite graph, $x, y :
    \Node_{G}$ and $n~:~\mathbb{N}$, the type $\mathsf{qswalk}(n,x,y)$
    in \Cref{def:finite-simple-walks} is a finite set. 
\end{lemma}
\begin{linkproof}[]{https://jonaprieto.github.io/synthetic-graph-theory/CPP2022-paper.html\#1777}
    It suffices to show that the type $\mathsf{qswalk}(n,x,y)$ is finite.
    The proof is by induction on $n$.
    \begin{enumerate} 
        \item If $n = 0$, the type defined by
        $\mathsf{qswalk}(0,x,z)$ is equivalent to the identity type $x=y$,
        as the only walks of length zero are the trivial walks. Given that
        the node set is discrete, the path space $x=y$ is
        finite by \Cref{lem:decidable-implies-finite-path}.
        \item Otherwise, given $x,z:\Node_{G}$, we must prove that the
            type $\mathsf{qswalk}(S(n),x,z)$ is finite, for $n :
            \mathbb{N}$, assuming that $\mathsf{qswalk}(n,x,z)$ is
            finite. This is equivalent to showing that the equivalent
            type given by \Cref{eq:equiv-type-fswalk} is finite. The
            required conclusion follows by
            \Cref{lem:finiteness-closure-property}, as each type of the
            $\Sigma$-type in the right-hand side of the equivalence in
            \Cref{eq:equiv-type-fswalk} is finite. The set $\Node_{G}$
            and the sets by $\Edge_{G}$ are each finite, as $G$ is a
            finite graph. The type $\mathsf{qswalk}(n, y, z)$ is finite
            by induction hypothesis. Lastly, any decidable proposition is finite
            i.e.  $(x \not \in w')$ is finite.\qedhere
    \end{enumerate}
    \end{linkproof}

\Cref{lem:lemma0,lem:lemma1} prove the fact mentioned earlier on the
node repetition condition in a quasi-simple walk. A node can only
appear once in a quasi-simple walk, unless the node is the end of the
walk. From now on, unless stated otherwise, we will refer to \(n\) as
the cardinality of \(\Node_{G}\) whenever the node set of the graph
\(G\) is finite. The number of nodes in any quasi-simple walk is
bounded by \(n+1\).

\begin{lemma}\label{lem:lemma1b} Let $G$ be a finite graph. Then
\Cref{eq:lemma1b} is a finite set.
\begin{equation}\label[type]{eq:lemma1b}
\sum_{(x,y~:~\Node_{G})}\sum_{(m~:~ ⟦ n + 1 ⟧ )} \mathsf{qswalk}(m,x,y).
\end{equation}
\end{lemma}

\begin{linkproof}[]{https://jonaprieto.github.io/synthetic-graph-theory/CPP2022-paper.html\#1803}
The conclusion follows since finite sets are closed under
$\Sigma$-types. $\Node_G$ is finite since $G$ is a finite graph. $⟦ n
+ 1 ⟧$ is finite. The type $\mathsf{qswalk}(m,x,y)$ is finite by
\Cref{lem:lemma2}.\qedhere
\end{linkproof}

\begin{lemma}\label{lem:lemma0} Given a graph $G$ with finite node set
of cardinality $n$, $x,y:\Node_{G}$ and a quasi-simple walk $w :
\mathsf{W}_{G}(x,y)$ of length $m$, then it holds that $m \leq n$.
\end{lemma}

\begin{linkproof}[]{https://jonaprieto.github.io/synthetic-graph-theory/CPP2022-paper.html\#1845}
It suffices to generate an embedding between the finite set $⟦m⟧$ and
the finite node set in $G$. Such an embedding is the projection
function $\pi_1~:~\Sigma_{x :\Node_{G}} (x \in w) \to \Node_{G}$.
Recall that the domain of the function $\pi_1$ is equivalent to $⟦m⟧$
by \Cref{lem:number-of-nodes-in-walk}.\qedhere
\end{linkproof}

Now, even when the type of walks forms an infinite set, thanks to
\Cref{lem:lemma0,thm:finite-simple-walks}, we will be able to prove
that for any nodes \(x\) and \(y\), the collection of quasi-simple
walks from \(x\) to \(y\) forms a finite set as long as the graph is
finite.

\begin{lemma}\label{lem:lemma1}
Given a graph $G$ with finite node set of cardinality $n$ and $x, y~:~\Node_{G}$,
the following equivalence holds.
\begin{equation}\label[equiv]{eq:lemma1}
\sum_{(w~:~\mathsf{W}_{G}(x,y))}  \mathsf{isQuasi}(w)
\simeq  \sum_{(m~:~⟦ n + 1 ⟧)} \mathsf{qswalk}(m,x,y).
\end{equation}
\end{lemma}

\begin{linkproof}[]{https://jonaprieto.github.io/synthetic-graph-theory/CPP2022-paper.html\#1883}
Apply \Cref{lem:lemma0}.\qedhere
\end{linkproof}

It is not immediately clear that quasi-simple walks forms a finite
set, even when the graph is finite. A quasi-simple walk can contain a
loop at its terminal node. One might think there are infinitely many
walks if each walk loops at its terminal nodes. However, it is by
constraining walks to be quasi-simple that we obtain the finiteness
property.

\begin{theorem}\label{thm:finite-simple-walks} 
The quasi-simple walks of a finite graph $G$ forms a finite set, i.e. 
\Cref{eq:finite-simple-walks} is inhabited.

\vspace{1mm}
\begin{equation}\label[type]{eq:finite-simple-walks}
\mathsf{isFinite}\left(\sum_{(x , y~:~\Node_{G})} \sum_{(w~:~\mathsf{W}(x,y))} \mathsf{isQuasi}(w)\right).
\end{equation}
\end{theorem}

\begin{linkproof}[]{https://jonaprieto.github.io/synthetic-graph-theory/CPP2022-paper.html\#1927}
The conclusion clearly follows from \Cref{lem:lemma1,lem:lemma1b}, since
finite sets are closed under type equivalences and $\Sigma$-types
by \Cref{lem:finiteness-closure-property}. \qedhere
\end{linkproof}

\hypertarget{walk-splitting}{%
\subsection{Walk Splitting}\label{walk-splitting}}

In this subsection, a function to split/divide a walk \(w\) from \(x\)
to \(z\) into subwalks, \(w_1\) and \(w_2\), is given. Such a division
of \(w\), of type \Cref{type:walk-division}, is handy e.g.~for proving
statements where the induction is not on the structure but on the
length of the walk.

\vspace{1mm}

\begin{equation}\label[type]{type:walk-division}
    \sum_{(y~:~\Node_{G})} \sum_{(w_1~:~\mathsf{W}_{G}(x,y))}\sum_{(w_2~:~\mathsf{W}_{G}(y,z))} (w = w_1 \cdot w_2).
\end{equation} \vspace{1mm}

Let \(x,y,z\) be variables for nodes in \(G\) and let \(w\) be a walk
from \(x\) to \(z\), unless stated otherwise. We refer to the walk
\(w_1\) in \Cref{type:walk-division} as a prefix of \(w\) and \(w_2\)
as the corresponding suffix given \(w_1\).

\begin{definition}\label{def:prefixes}
Given two walks $p$ and $q$ with the same head, one says that $p$ is a
\emph{prefix} of $q$ if the type $\mathsf{Prefix}(p,q)$ is inhabited.

\vspace{1mm}
\begin{equation*}
    \begin{aligned}
\mathsf{\textbf{data}} & \; \mathsf{Prefix}~: \Pi\,\{x,y,z\}\,.\,\mathsf{W}_{G}(x,y) \to \mathsf{W}_{G}(x,z) → \UU \; \mathsf{ } \\
& \mathsf{head}~:~\Pi\,\{x\,y\}\, .\,\Pi\,\{w~:~\mathsf{W}_{G}(x,y)\}\,.\,\mathsf{Prefix}(⟨x⟩,w)\\
& \mathsf{by\mbox{-}edge}
: \Pi\,\{x\,y\,z\,k\}\,.\,\Pi\,\{e~:~\Edge_{G}(x,y)\} \\
&\hspace{12.5mm} .\ \Pi\,\{p~:~\mathsf{W}_{G}(y,z)\}\,.\,\Pi\,\{q~:~\mathsf{W}_{G}(y,k)\} \\
&\hspace{12mm} \to \mathsf{Prefix}(p, q) \to \mathsf{Prefix}(e ⊙ p, e ⊙ q)\\
    \end{aligned}
\end{equation*}
\vspace{.5mm}

\end{definition}

\begin{lemma}\label{lem:find-suffix} Given a prefix $w_1$ for a walk
$w$, we can prove that there is a term of type \Cref{type:suffix}
named $\textsf{suffix}(w_1, w, t)$, referring to as the suffix of $w$
given $w_1$, where $t~:~w = w_1 \cdot w_2$.

\begin{equation}\label[type]{type:suffix} 
    \sum_{(w_2~:~\mathsf{W}_{G}(y,z))}\, (w = w_1 \cdot w_2) .
\end{equation}

\end{lemma}

\begin{linkproof}[]{https://jonaprieto.github.io/synthetic-graph-theory/CPP2022-paper.html\#2022}
For brevity, we skip the trivial cases for $w_1$ and $w$. The
remaining cases are proved by induction; first, on $w_1$, and secondly
on $w$. The resulting nontrivial case occurs when $w_1 = e ⊙ p$, $w =
e ⊙ q$ and $t:\mathsf{Prefix}(p,q)$ for two walks $p$ and $q$. By
the induction hypothesis applied to $p,q$, and $t$, the term
$\mathsf{suffix}(p,q,t)$ is obtained, from which one gets the suffix
walk $w_2$ along with a proof $i~:~q = p \cdot w_2$. Thus, the
required term is the pair $(w_2, \mathsf{ap}(e ⊙\mbox{-},i))$.\qedhere
\end{linkproof}

We now encode the case where the walk \(w\) is divided at the first
occurrence of the node \(y\), using the type family
\(\mathsf{SplitAt}(w,y)\) defined in \Cref{def:type-splitat}. The
corresponding method to inhabit the type \(\mathsf{SplitAt}(w, y)\) is
the function given in \Cref{def:view-splitat}, assuming the node set
in the graph is discrete. This walk splitting encoding is implicitly
used in several parts of the proof of \Cref{thm:hom-normalisation}.

\begin{definition}\label{def:type-splitat} The type
$\mathsf{SplitAt}(w, y)$ is the inductive type defined as:
\vspace{.5mm}
\begin{equation*}
    \begin{aligned}
\mathsf{\textbf{data}} & \; \mathsf{SplitAt}\ \{x\,z\} (w~:~\mathsf{W}_{G}(x,z))\,(y~:~\Node_{G}) :\; \UU \; \mathsf{ } \\
& \mathsf{nothing}~:~\Pi\,\{x\,y\}\,.\,\Pi\,\{w~:~\mathsf{W}_{G}(x,y)\}\\
&\hspace{12mm} \to (y \not \in w)\\
&\hspace{12mm} \to \mathsf{SplitAt}(w,y)\\
& \mathsf{just} : \Pi\,\{x\,y\}\,.\,\Pi\,\{w~:~\mathsf{W}_{G}(x,y)\}\\
&\hspace{5.5mm} \to (p : \mathsf{W}_{G}(x,y)) \\
&\hspace{5.5mm} \to \mathsf{Prefix}(p, w) \to (y \not \in p)\\
&\hspace{5.5mm}\to \mathsf{SplitAt}(w,y)\\
\end{aligned}
\end{equation*}
\vspace{.5mm}
\end{definition}

\begin{lemma}\label{def:view-splitat} The type $\mathsf{SplitAt}(w,y)$
    is inhabited if the node set of the graph is discrete.
\end{lemma}

\begin{linkproof}[]{https://jonaprieto.github.io/synthetic-graph-theory/CPP2022-paper.html\#2064}
By induction on the structure of the walk.
\begin{enumerate}
    \item If the walk is trivial, then the required term is
    $\mathsf{nothing} \mathsf{id}$, as by definition, $y \not \in \mathbb{0}$.
    \item If the walk is the composite $(e ⊙ w)$ with $e :
    \Edge_{G}(x,y')$ and $w~:~\mathsf{W}_{G}(y',z)$, we ask whether
    $y$ is equal to $x$ or not.
    \begin{enumerate}
        \item If $y = x$ then the required term is
        $\mathsf{just}(\langle y
        \rangle,\mathsf{head},\mathsf{id})$.
        \item If $y \neq x$ then by the induction hypothesis on $w$ and
          $y$, the following cases need to be considered.
        \begin{enumerate}

            \item If the case is $\mathsf{nothing}$, then there is enough evidence
            that $y\not \in w$ and we use for the required term
            the $\mathsf{nothing}$ constructor.

            \item Otherwise, there is a prefix $w_1$ for $w$ and a
            proof $r : y \not \in w_1$. Using $r$ and the fact $x \neq
            y$, we can construct $r' : y \not \in (e ⊙ w_1)$. Then,
            the term that we are looking for is $\mathsf{just}(e ⊙
            w_1, \mathsf{by\mbox{-}edge}(p), r')$ of type
            $\mathsf{SplitAt}(e ⊙ w, y)$, as required in the
            conclusion. \qedhere
        \end{enumerate}
    \end{enumerate}
\end{enumerate}
\end{linkproof}

\subsection{Normal Forms for Walks}\label{sec:loop-reduction-relation}

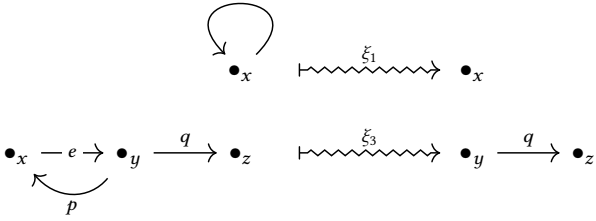
\begin{figure}
\centering
\begin{tikzcd}
& {} & \bullet_{x} &[-2em] {} &[3em] {} &[-3em]\bullet_{x} &{} &{}\\
\bullet_{x} &[2em] \bullet_{y} &  \bullet_{z} & {} &{} &\bullet_{y} &\bullet_{z}
\arrow[from=1-3, to=1-3, loop]
\arrow["e"{description}, from=2-1, to=2-2]
\arrow["p", curve={height=-18pt}, from=2-2, to=2-1]
\arrow["q", from=2-2, to=2-3]
\arrow["q", from=2-6, to=2-7]
\arrow["{\xi_1}",  maps to, squiggly, from=1-4, to=1-5]
\arrow["{\xi_3}",  maps to, squiggly, from=2-4, to=2-5]
\end{tikzcd}
\caption{The rules $\xi_1$ and $\xi_3$ of the loop-reduction relation
in \Cref{eq:reduction-relation}.}
\label{fig:loop-reduction}
\end{figure}

In this subsection, a reduction relation in
\Cref{def:loop-reduction-relation} is established on the set of walks
of equal endpoints. Some cases considered by such a relation are
illustrated in \Cref{fig:loop-reduction}. This relation provides a way
to remove loops from walks in a graph with a discrete set of nodes.
The notion of normal form for walks presented in this work is based on
the loop reduction relation in \Cref{def:normal-form}.

The following definitions establish a few type families to encode
walks of a certain basic structure---for example, nontrivial walks and
loops---necessary for the formalisation.

\begin{definition}
Let $x,y~:~\Node_{G}$ and $w :\mathsf{W}_{G}(x,y)$.
\begin{enumerate}
\item The walk $w$ is a loop whenever the head is equal to the end, i.e.
$\mathsf{Loop}(w)$.
\begin{equation*}
\begin{aligned}
\mathsf{\textbf{data}} & \; \mathsf{Loop}~:~\, \Pi\,\{x,y\}\,.\,\mathsf{W}_{G}(x,y) → \UU \; \mathsf{ } \\
& \mathsf{is\mbox{-}loop}: \,\Pi \{x\,y\}.\,\Pi\{w~:~\mathsf{W}_{G}(x,y)\}\\
&\hspace{10mm} → x = y \to \mathsf{Loop}(w)
\end{aligned}
\end{equation*}

\item The walk $w$ is trivial if its length is zero, i.e.  $\mathsf{Trivial}(w)$.%
\begin{equation*}
\begin{aligned}
\mathsf{\textbf{data}} & \; \mathsf{Trivial}~:~\, \Pi\,\{x,y\}\,.\,\mathsf{W}_{G}(x,y) → \UU \; \mathsf{ } \\
& \mathsf{is\mbox{-}trivial}: \,\Pi \{x\,y\}.\,\Pi\{w : \mathsf{W}_{G}(x,y)\}\\
&\hspace{12.5mm} → \mathsf{length}(w) = 0 \to \mathsf{Trivial}(w)
\end{aligned}
\end{equation*}
\item A walk $w$ is not trivial, if it has one edge at least, i.e.
$\mathsf{NonTrivial}(w)$.
\begin{equation*}
\begin{aligned}
\mathsf{\textbf{data}} & \; \mathsf{NonTrivial} : \, \Pi\,\{x,y\}\,.\,\mathsf{W}_{G}(x,y) → \UU \; \mathsf{ } \\
& \mathsf{has\mbox{-}edge}\,:\,\Pi \{x\,y\,z\}.\,\Pi\{w : \mathsf{W}_{G}(y,z)\}\\
&\hspace{14mm}\to(e : \Edge_{G}(x,y)) → \mathsf{NonTrivial}(e ⊙ w).
\end{aligned}
\end{equation*}
\item A walk $w$ does not \emph{reduce} if $\mathsf{NoReduce}(w)$.
\begin{equation*}
    \begin{aligned}
\mathsf{\textbf{data}} & \; \mathsf{NoReduce} : \, \Pi\,\{x,y\}\,.\,\mathsf{W}_{G}(x,y) → \UU \; \mathsf{ } \\
& \mathsf{is}\mbox{-}\mathsf{dot} : \Pi\{x\}\,.\,\mathsf{NoReduce}(⟨ x ⟩) \\
& \mathsf{is}\mbox{-}\mathsf{edge} : \Pi\{x\,y\}.\,\Pi\,\{e : \Edge_{G}(x,y)\} \\
&\hspace{10.5mm} → (x ≠ y) → \mathsf{NoReduce}(e ⊙ ⟨ y ⟩)
    \end{aligned}
\end{equation*}
\item A walk $w$ is not a trivial loop if $\mathsf{NonTrivialLoop}(w)$.
\end{enumerate}
\begin{equation*}
    \begin{aligned}
\mathsf{\textbf{data}} & \; \mathsf{NonTrivialLoop} : \, \Pi\,\{x,y\}\,.\,\mathsf{W}_{G}(x,y) → \UU \; \mathsf{ } \\
&\mathsf{is}\mbox{-}\mathsf{loop} : \Pi\{x\,y\,z\}\,.\{e : \Edge_{G}(x,y)\}\\
&\hspace{10mm} → (p : x = z )\,→\,(w : \mathsf{W}_{G})\\ 
&\hspace{10mm} → \,\mathsf{NonTrivialLoop}(e ⊙ w)
  \end{aligned}
\end{equation*}
\end{definition}

\begin{lemma} Given $x,y : \Node_{G}$ and $u: \mathsf{W}_{G}(x,y)$, the
following claims hold.
\begin{enumerate}
    \item If $x \neq y$ then $\mathsf{NonTrivial}(u)$.
    \item If $\mathsf{NonTrivial}(u)$ then $x \in u$.
    \item Given $z : \Node_{G}$, if $\mathsf{NonTrivial}(u)$ and
    $v : \mathsf{W}_{G}(y,z)$ then $\mathsf{NonTrivial}(u \cdot v)$.
\end{enumerate}
\end{lemma}

\noindent Remember that a reduction relation \(R\) on a set \(M\) is
an irreflexive binary relation on \(M\). If \(R\) is a reduction
relation, we use \(xRy\) to refer to the pair \((x,y)\) in \(R\). If
\(xRy\) then one says that \(x\) \emph{reduces} to \(y\) or simply
\(x\) \emph{reduces}.

\begin{definition}\label{def:loop-reduction-relation} The
\href{https://jonaprieto.github.io/synthetic-graph-theory/CPP2022-paper.html\#2358}{\emph{loop-reduction}
relation} ($\rightsquigarrow$) on walks is
\Cref{eq:reduction-relation}.

\begin{equation}\label[type]{eq:reduction-relation}
    \begin{aligned}
    \mathsf{\textbf{data}} & \;(\rightsquigarrow)  : \, \Pi\,\{x,y : \Node_{G}\}.\mathsf{W}_{G}(x,y) \to \mathsf{W}_{G}(x,y)  → \UU \; \mathsf{ } \\
    & ξ₁ : \Pi\,\{x\,y\}\,.\,(p : \mathsf{W}_{G}(x,y))\,(q : \mathsf{W}_{G}(x,y)) \\
    &\hspace{3mm} → \mathsf{NonTrivialLoop}(p) → \mathsf{Trivial}(q)
    \\
          &\hspace{3mm} → p \rightsquigarrow q \\
    & ξ₂ : \Pi\,\{x\,y\,z\}\,.\,(e : \Edge_{G}(x,y))\,(p, q : \mathsf{W}_{G}(y,z))
    \\
    &\hspace{3mm} → ¬\,\mathsf{Loop}(e ⊙ p)  → x ≠ y\\
          &\hspace{3mm}
          → (p \rightsquigarrow q)
          → (e ⊙ p) \rightsquigarrow (e ⊙ q)
        \\
    & ξ₃ : \Pi\,\{x\,y\,z\}\,.\, (e : \Edge_{G}(x,y))\,(p : \mathsf{W}_{G}(y,x))\\
        &\hspace{3mm} → (q : \mathsf{W}_{G}(x,z)) \\
          &\hspace{3mm} → ¬\,\mathsf{Loop} ((e ⊙ p) \cdot q)
          → \mathsf{Loop} (e ⊙ p)\\
          &\hspace{3mm} → \mathsf{NonTrivial}(q)
          \\
          &\hspace{3mm} → (w : \mathsf{W}_{G}(x,z))
          → w = (e ⊙ p) \cdot q \\
          &\hspace{3mm}
          → w \rightsquigarrow q
    \end{aligned}
    \end{equation}

The following provides hints to the intuition behind each of the data
constructors above.

\begin{enumerate}
\item The rule ξ₁ is \say{a nontrivial loop reduces to the trivial
walk of its endpoint}.
\item The rule ξ₂ is \say{the relation ($\rightsquigarrow$) is right
compatible with edge concatenation}.
\item The rule ξ₃ is \say{the relation ($\rightsquigarrow$) removes
right attached loops}.
\end{enumerate}

\end{definition}

\begin{remark} The data constructors in \Cref{eq:reduction-relation}
follow a design principle to avoid certain unification problems
occurring in dependently type programs \citep{conorgreenslime, plfa}. 
\end{remark}

\begin{definition}
The relation $(\rightsquigarrow^{*})$ is the reflexive and transitive
closure of the relation $(\rightsquigarrow)$ in
\Cref{def:loop-reduction-relation}.
\end{definition}

\begin{lemma}\label{lem:nf} Given $x,y : \Node_{G}$ and  $p,q :
\mathsf{W}_{G}(x,y)$, the following claims hold:
\begin{enumerate}
    \item\label{lem:nf-positive} If $x ∈ q$ and  $p
    \rightsquigarrow^{*} q$ then $x ∈ p$.
    \item\label{lem:nf-length} If $p \rightsquigarrow q$ then
    $\mathsf{length}(q) < \mathsf{length}(p)$.
\end{enumerate}
\end{lemma}

One can prove that our reduction relation in
\Cref{def:loop-reduction-relation} satisfies the progress property,
similarly as proved for simply-typed lambda calculus in Agda \citep[
§2]{plfa}. The evidence that a walk reduces is encoded using the
following predicate.

\begin{definition}\label{def:Reduce}
 Given a walk $p : \mathsf{W}_{G}(x,y)$,
\hspace*{5mm}
\begin{equation*}
\mathsf{Reduce}(p) :\equiv \sum_{(q~:~\mathsf{W}_{G}(x,y))} (p \rightsquigarrow q).
\end{equation*}
\end{definition}

The predicate \(\mathsf{Normal}\) defined in \Cref{def:normal-form} is
the evidence that a walk is a quasi-simple walk that can no longer
reduce.

\begin{definition}\label{def:normal-form} Given a walk $p$, one states
that $p$ is in \emph{normal form} if $\mathsf{Normal}(p)$. If
$p\rightsquigarrow q$ and $q$ is in normal form, we refer to $q$ as the
normal formal of $p$.
\begin{equation*}\label{eq:normal-form}
    \mathsf{Normal}(p) :\equiv \mathsf{isQuasi}(p) \times ¬\,\mathsf{Reduce}(p).
\end{equation*}
\end{definition}

\begin{lemma}
Being in normal form for a walk is a proposition.
\end{lemma}
\begin{linkproof}[]{https://jonaprieto.github.io/synthetic-graph-theory/CPP2022-paper.html\#2479}
It follows from \Cref{lem:finiteness-closure-property,lem:being-simple-is-prop}.\qedhere
\end{linkproof}

\begin{example} \label{No-reduce-no-step} The very basic normal forms
    for walks are the trivial ones and the one-edge walks with
    different endpoints. Given a walk $w$ and a term of
    $\mathsf{NoReduce}(w)$, one can easily show that the walk $w$ is
    in normal form. 
\end{example}

\begin{definition}\label{def:progress} Given nodes $x$ and $y$ in a graph
$G$, we encode the fact a walk can reduce or not by using the inductive data
type $\mathsf{Progress}$.\\[2mm]
\begin{equation*}
    \begin{aligned}
        \mathsf{\textbf{data}} & \; \mathsf{Progress}\,\{x\,y\}\;(p : \mathsf{W}_{G}(x,y))  : \, \UU \; \mathsf{ } \\
        & \mathsf{step}\,:\,\mathsf{Reduce}(p) → \mathsf{Progress}(p)\\
        & \mathsf{done}\,:\,\mathsf{Normal}(p) → \mathsf{Progress}(p)
    \end{aligned}
\end{equation*}
\vspace{.5mm}

\end{definition}

\begin{theorem}
    \label{thm:normalisation} 
    Given a graph $G$ with a discrete node set, there
     exists a reduction for each walk to one of its normal forms, i.e. 
     \Cref{eq:predicate-P} is inhabited for all $w :
     \mathsf{W}_{G}(x,y)$.
    \begin{equation}\label[type]{eq:predicate-P}
        \sum_{(v~:~\mathsf{W}_{G}(x,z))} (w \rightsquigarrow^{*} v) × \mathsf{Normal}(v).
    \end{equation}\vspace*{-2mm}
\end{theorem}

\begin{remark} The reduction relation $(\rightsquigarrow)$ has the
termination property. There is no infinite sequence of walks reducing,
since the length of each walk in a chain like $w_1\rightsquigarrow
w_2\rightsquigarrow w_3\rightsquigarrow \cdots$, decreases at each
reduction step. See also \Cref{lem:well-founded-walk-relation}.
\end{remark}

\begin{corollary}
    \label{thm:progress} Given a graph $G$ with a discrete node set,
and a walk $w$ of type $\mathsf{W}_{G}(x,y)$ for two $x,y :
\Node_{G}$, the following claims hold.
\begin{enumerate}
    \item The type $\mathsf{Reduce}(w)$ is decidable.
    \item The proposition $\mathsf{Normal}(w)$ is decidable.
    \item The walk $w$ progresses in the sense of \Cref{def:progress}.
\end{enumerate}
\end{corollary}

For simplicity, the proofs of \Cref{thm:normalisation,thm:progress}
are omitted. Neither of them requires the law of excluded middle.
However, if we want to construct the normal form for a walk, the node
set of the graph has to be discrete. In the case of
\Cref{thm:normalisation}, its proof can use the same reasoning given
for the proof of \Cref{thm:hom-normalisation}.

\section{The Notion of Walk Homotopy}\label{sec:homotopy-normalisation}

\begin{figure*}[!htb]
\includegraphics[width=0.9\textwidth]{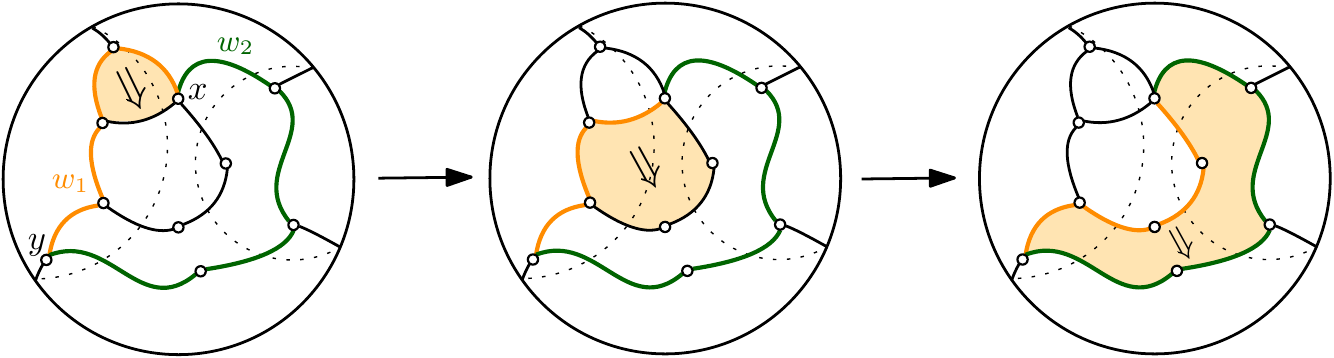}
\caption{It is shown three homotopies between two walks from $x$ to
$y$ in a graph embedded in the sphere. In each case, the arrow
$(\Downarrow)$ indicates the face and the direction in which the
corresponding walk deformation is performed. We obtain a homotopy
between the two highlighted walks, $w_1$ and $w_2$, by composing, from
left to right, the homotopies from each figure.}
\label{fig:walk-homotopies}
\end{figure*}

This section introduces the notion of homotopy for walks denoted by
\((\sim_{\mathcal{M}})\). We define such a relation in
\Cref{def:congruence-relation} as a congruence relation on the
category induced by the endofunctor (W) on the corresponding graph.
Because homotopy for walks depends on the surface in which the graph
is embedded, it is necessary to first define an embedding of graphs in
a surface.

A map/embedding of a graph is a cellular decomposition of the surface
where the graph is embedded. This topological definition also requires
defining what a surface is. To avoid this, we consider instead a
combinatorial approach in \Cref{def:graph-map} based on the work by
Edmonds and Tutte \citep{Tutte1960, Tutte1963}. A more complete
description of graph maps can be found in \citep[§3]{gross}.

Given a graph \(G\), the graph formed by taking the same node set of
\(G\) and the edge set as the type \(\Edge_{G}(x,y) + \Edge_{G}(y,x)\)
for \(x,y:\Node_{G}\) is denoted by \(U(G)\) and referred as the
\emph{symmetrisation} of \(G\).

\begin{definition}\label{def:graph-map} A map for a graph $G$
  of type $\mathsf{Map}(G)$ is a local \emph{rotation system} at each
  node in $U(G)$.
\begin{align*}
  \mathsf{Map}(G)  &:≡ \prod_{(x~:~\Node_{G})} \mathsf{Cyclic}\left(\sum_{(y~:~\Node_{G})}\Edge_{U(G)}(x,y) \right).
\end{align*}
\end{definition}

Given a map \(\mathcal{M}\), the \emph{faces} of \(\mathcal{M}\) are
the regions obtained by the cellular decomposition of the
corresponding surface by \(\mathcal{M}\). We omit
\href{https://jonaprieto.github.io/synthetic-graph-theory/lib.graph-embeddings.Map.Face.html\#2355}{the
formal type} of faces herein, so as not to distract the reader from
the goals of this paper. The type of faces requires proper attention
\citep{planarityHoTT}. Put briefly, a face is a cyclic walk in the
embedded graph without repeating nodes and without edges inside
\citep{gross}. The corresponding data of a face is a cyclic subgraph
\(A\) in \(U(G)\) and a function \(f : A \to N_{G}\) that picks nodes
in \(A\). Consequently, for each face \(\mathcal{F}\) given by
\(\langle A, f\rangle\), there are at least two quasi-simple walks in
\(U(G)\) associated with \(\mathcal{F}\) for every node-pair. Given
\(x,y:\Node_{G}\), the corresponding walks given by \(\mathcal{F}\)
are, namely, the clockwise and counter-clockwise closed walks in
\(U(G)\), denoted by \(\mathsf{cw}_{A}(x,y)\) and
\(\mathsf{ccw}_{A}(x,y)\), respectively. If the endpoints are equal,
the trivial walk \(\langle x\rangle\) must also be considered.

\begin{figure}[!htb]
    \centering
\[\begin{tikzcd}[column sep=normal]
  {\bullet_{x}} & {\bullet_{f(a)}} && {\bullet_{f(b)}} & {\bullet_{y}}
  \arrow["{w_1}", from=1-1, to=1-2]
  \arrow[""{name=0, anchor=center, inner sep=0}, "{\mathsf{ccw}_\mathcal{F}(a,b)}"', curve={height=18pt}, from=1-2, to=1-4]
  \arrow["{w_2}", from=1-4, to=1-5]
  \arrow[""{name=1, anchor=center, inner sep=0}, "{\mathsf{cw}_\mathcal{F}(a,b)}", curve={height=-18pt}, from=1-2, to=1-4]
  \arrow[shorten <=5pt, shorten >=5pt, Rightarrow, from=1, to=0]
\end{tikzcd}\] \caption{Given a face $\mathcal{F}$ of the map
  $\mathcal{M}$, we illustrate here $\mathsf{hcollapse}$, one of the
  four constructors of the homotopy relation on walks in
  \Cref{def:congruence-relation}. The arrow $(\Downarrow)$ represents a
  homotopy of walks.}
  \label{fig:constructors-for-homotopic-walks}
  \end{figure}
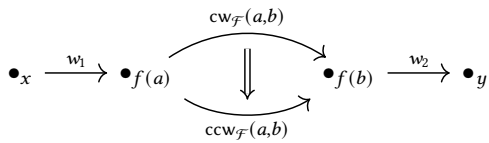

\subsection{Homotopy of Walks}\label{sec:homotopy-of-walks}

\begin{definition}\label{def:congruence-relation} Let $w₁,w₂$ be two
walks from $x$ to $y$ in $U(G)$. The expression
$\HomWalk{\mathcal{M}}{w₁}{w₂}$ denotes that one can \emph{deform}
$w_1$ into $w_2$ along the faces of $\mathcal{M}$, as illustrated in
\Cref{fig:walk-homotopies}. We acknowledge the evidence of this
deformation as a walk homotopy between $w_1$ and $w_2$, of type
$\HomWalk{\mathcal{M}}{w₁}{w₂}$. The relation $(\sim_{\mathcal{M}})$
has four constructors as follows. The first three constructors are
functions to indicate that homotopy for walks is an equivalence
relation, they are $\mathsf{hrefl}$, $\mathsf{hsym}$, and
$\mathsf{htrans}$. The fourth constructor, illustrated in
\Cref{fig:constructors-for-homotopic-walks}, is the
$\mathsf{hcollapse}$ function that establishes the walk homotopy:

\begin{equation*}\label{eq:collapse}
    \HomWalk{\mathcal{M}}{(w₁ \cdot
\mathsf{ccw}_{\mathcal{F}}(a,b) \cdot w₂)} {(w₁ \cdot
\mathsf{cw}_{\mathcal{F}}(a,b) \cdot w₂)},
\end{equation*}

supposing one has the following,
\begin{itemize}
  \item[(i)] a face $\mathcal{F}$ given by $\langle A, f \rangle$ of the
  map $\mathcal{M}$,
  \item[(ii)] a walk $w₁$ of type $\mathsf{W}_{U(G)}(x,f(a))$ for a node $x$ in $G$
  with a node $a$ in $A$, and
  \item[(iii)] a walk $w₂$ of type $\mathsf{W}_{U(G)}(f(b),y)$ for a node $b$ in $A$
  with a node $y$ in $G$.
\end{itemize}
\end{definition}

The following lemma shows how to compose walk homotopies horizontally
and vertically. We consider a map \(\mathcal{M}\) for a graph \(G\)
and distinguishable nodes, \(x,y\), and \(z\) where \(w\), \(w_1\),
and \(w_2\) are walks from \(x\) to \(y\).

\begin{lemma} \label{lem:whiskering}
  \hspace*{5cm}
\begin{enumerate}
  \item (Right whiskering) Let $w_3$ be a walk of type
    $\mathsf{W}_{U(G)}(y,z)$. If $\HomWalk{\mathcal{M}}{w₁}{w₂}$ then $\HomWalk{\mathcal{M}}{(w₁
    \cdot w₃)}{(w₂ \cdot w₃)}$.
\begin{center}
\begin{tikzcd}
  \bullet_{x} & \bullet_{y} & \bullet_{z} & [-5mm]{\color{darkblue} \to} &[-3mm]\bullet_{x} & \bullet_{z}
  \arrow[""{name=0, anchor=center, inner sep=0}, "{w_1}", curve={height=-12pt}, from=1-1, to=1-2]
  \arrow[""{name=1, anchor=center, inner sep=0}, "{w_2}"', curve={height=12pt}, from=1-1, to=1-2]
  \arrow["{w_3}", from=1-2, to=1-3]
  \arrow[""{name=2, anchor=center, inner sep=0}, "{w_1 \cdot w_3}", curve={height=-12pt}, from=1-5, to=1-6]
  \arrow[""{name=3, anchor=center, inner sep=0}, "{w_2 \cdot w_3}"', curve={height=12pt}, from=1-5, to=1-6]
  \arrow[shorten <=3pt, shorten >=3pt, Rightarrow, from=2, to=3]
  \arrow[shorten <=3pt, shorten >=3pt, Rightarrow, from=0, to=1]
\end{tikzcd}
\end{center}

\item (Left whiskering) Let $p₁,p₂$ be walks of type
  $\mathsf{W}_{U(G)}(y,z)$. If $\HomWalk{\mathcal{M}}{p₁}{p₂}$ then $\HomWalk{\mathcal{M}}{(w
  \cdot p₁)}{(w \cdot p₂)}$.
  \begin{center}
    \begin{tikzcd}
  \bullet_{x} & \bullet_{y} & \bullet_{z} &[-5mm]{\color{darkblue} \to} &[-3mm]\bullet_{x} & \bullet_{z}
  \arrow["w", from=1-1, to=1-2]
  \arrow[""{name=0, anchor=center, inner sep=0}, "{p_1}", curve={height=-12pt}, from=1-2, to=1-3]
  \arrow[""{name=1, anchor=center, inner sep=0}, "{w\cdot p_1}", curve={height=-12pt}, from=1-5, to=1-6]
  \arrow[""{name=2, anchor=center, inner sep=0}, "{w \cdot p_2}"', curve={height=12pt}, from=1-5, to=1-6]
  \arrow[""{name=3, anchor=center, inner sep=0}, "{p_2}"', curve={height=12pt}, from=1-2, to=1-3]
  \arrow[shorten <=3pt, shorten >=3pt, Rightarrow, from=1, to=2]
  \arrow[shorten <=3pt, shorten >=3pt, Rightarrow, from=0, to=3]
  \end{tikzcd}
\end{center}
\item (Full whiskering) Let $p₁,p₂$ be walks of type
  $\mathsf{W}_{U(G)}(y,z)$. If $\HomWalk{\mathcal{M}}{w₁}{w₂}$ and
  $\HomWalk{\mathcal{M}}{p₁}{p₂}$, then $\HomWalk{\mathcal{M}}{(w₁ \cdot p₁)}{(w₂ \cdot
  p₂)}$.
\begin{center}
\begin{tikzcd}
  \bullet_{x} & \bullet_{y} & \bullet_{z} & [-5mm]{\color{darkblue} \to} &[-3mm]\bullet_{x} & \bullet_{z}
  \arrow[""{name=0, anchor=center, inner sep=0}, "{w_1}", curve={height=-12pt}, from=1-1, to=1-2]
  \arrow[""{name=1, anchor=center, inner sep=0}, "{w_2}"', curve={height=12pt}, from=1-1, to=1-2]
  \arrow[""{name=2, anchor=center, inner sep=0}, "{p_1}", curve={height=-12pt}, from=1-2, to=1-3]
  \arrow[""{name=3, anchor=center, inner sep=0}, "{p_2}"', curve={height=12pt}, from=1-2, to=1-3]
  \arrow[""{name=4, anchor=center, inner sep=0}, "{w_1 \cdot p_1}", curve={height=-12pt}, from=1-5, to=1-6]
  \arrow[""{name=5, anchor=center, inner sep=0}, "{w_2 \cdot p_2}"', curve={height=12pt}, from=1-5, to=1-6]
  \arrow[shorten <=3pt, shorten >=3pt, Rightarrow, from=2, to=3]
  \arrow[shorten <=3pt, shorten >=3pt, Rightarrow, from=4, to=5]
  \arrow[shorten <=3pt, shorten >=3pt, Rightarrow, from=0, to=1]
\end{tikzcd}
\end{center}
\end{enumerate}
\end{lemma}

\subsection{Homotopy Walks in the Sphere}\label{sec:homotopy-walks-in-sphere}

In topology, the property of being simply connected to the sphere
states that one can freely deform/contract any walk on the sphere into
another whenever they share the same endpoints. This topological
property of the sphere motivates the predicate in
\Cref{def:spherical-map}, which establishes the conditions necessary
for embedding a graph into a sphere. Later, we show an alternative
definition for graphs with a node set in
\Cref{def:spherical-map-simple}. Given a distinguished face in a
connected graph, being spherical for a graph embedding serves to
establish elementary planarity criteria for graphs
\citep{planarityHoTT}.

\begin{definition}\label{def:spherical-map} Given a graph $G$, a map
$\mathcal{M}$ for $G$ is \emph{traditionally} spherical if
\Cref{eq:traditional-spherical} is inhabited.

\begin{equation}\label[type]{eq:traditional-spherical}
\prod_{(x,y~:~\Node_{G})} \prod_{(w₁, w₂~:~\mathsf{W}_{U(G)}(x,y))}
∥\HomWalk{\mathcal{M}}{w₁}{w₂} ∥.
\end{equation}
\end{definition}

To prove a given map is spherical following \Cref{def:spherical-map},
one must consider the set of all possible walk-pairs for each
node-pair. This is not easy, unless the set of walks follows a certain
property, since the type of walks forms an infinite set. Therefore, it
is proposed an alternative formulation for spherical maps based on
\Cref{def:loop-reduction-relation}. Any walk is homotopic to its
normal form, and only quasi-simple walks can be in normal form. By
removing such a \say{redundancy} created by loops in the graph, a more
convenient definition is obtained for spherical maps for graphs with
discrete node set, see \Cref{def:spherical-map-simple}. Furthermore,
using \Cref{thm:hom-normalisation}, we show that both definitions are
equivalent for graphs with discrete node set in
\Cref{lem:two-spherical-map-definition-are-equivalent}.

\begin{definition}\label{def:spherical-map-simple} Given a graph $G$,
a map $\mathcal{M}$ for $G$ is spherical if the type
\Cref{eq:quasi-spherical} is inhabited.

\begin{equation}\label[type]{eq:quasi-spherical}
\begin{split}
\prod_{(x,y~:~\Node_{G}}\prod_{(w₁, w₂~:~\mathsf{W}_{U(G)}(x,y))}\,
\mathsf{isQuasi}(w₁)\,&\times\,\mathsf{isQuasi}(w₂)\\[-5mm]
&\to\, ∥ \HomWalk{\mathcal{M}}{w₁}{w₂} ∥. 
\end{split}
\end{equation} 
\end{definition}

We will only refer to spherical maps as maps that follow
\Cref{def:spherical-map-simple}, unless stated otherwise. It is
straightforward to prove that loops are homotopic to the corresponding
trivial walk if a spherical map is given.

\begin{lemma}
\label{lem:loop-edges-homotopic-to-trivial-walks} Given a
    graph $G$, a spherical map $\mathcal{M}$ and $x : \Node_{G}$, it
    follows that $\| (e ⊙ ⟨ x ⟩)  \sim_{\mathcal{M}} ⟨x⟩ \|$ for all
    $e : \Edge_{U(G)}(x,x)$.
\end{lemma}
\begin{linkproof}[]{https://jonaprieto.github.io/synthetic-graph-theory/CPP2022-paper.html\#2923}
Apply  $\mathcal{M}$ to the walks $(e ⊙ ⟨ x ⟩)$ and $⟨x⟩$.\qedhere
\end{linkproof}

\begin{theorem}
\label{thm:hom-normalisation} Given a graph $G$ with a
    spherical map $\mathcal{M}$ and discrete set of nodes, for any
    walk $p : \mathsf{W}_{U(G)}(x,z)$, there exists a normal form of
    $p$, denoted by $\mathsf{nf}(p)$, such that $p$ is merely
    homotopic to $\mathsf{nf}(p)$, in the sense of
    \Cref{def:congruence-relation}. \end{theorem}

\begin{linkproof}[]{https://jonaprieto.github.io/synthetic-graph-theory/CPP2022-paper.html\#2957}
\label{proof:thm:hom-normalisation}
Given a walk $p$ in $U(G)$ from $x$ to $z$ of length $n$, we will
construct a term of type $Q(\mathcal{M},x,z,p)$ defined as follows.%
\begin{equation*}\label{eq:predicate-Q}
  Q(\mathcal{M}, x,z,w) :\equiv \hspace{-7mm}\sum_{(v~:~\mathsf{W}_{U(G)}(x,z))}
  \hspace{-3mm}(w \rightsquigarrow^{*} v)\,\times\,\mathsf{Normal}(v)\,\times\,\| w\sim_{\mathcal{M}}v \|.
\end{equation*}%
The proof is done by using strong induction on $n$. 
\begin{itemize}
    \item Case $n$ equals zero. The walk $p$ is the trivial walk
    $\langle x \rangle$, and it is then in normal form and also, by
    $\mathsf{hrefl}$, homotopic to itself.
    \item Case $n$ equals one. The walk $p$ is a one-edge walk. We
    then ask if $x = z$.
    \begin{enumerate}
        \item If $x = z$, the walk $p$ reduces to the trivial walk
        $\langle x\rangle$ by $\xi_1$. Applying $\mathcal{M}$, one
        obtains evidence of a homotopy between $p$ and $\langle x
        \rangle$, as the two walks are quasi-simple.
        \item If $x \neq z$, the one-edge walk $p$ is its own normal form
        and homotopic to itself by $\mathsf{hrefl}$.
    \end{enumerate}
    \item Assuming that $Q(x', z', w)$ for any walk $w$ from $x'$ to
    $y'$ of length $k \leq n$, we must prove that $Q(x,z,p)$ when the
    length of $p$ is $n + 1$.
    \item Therefore, let $p$ be a walk $(e~⊙~w)$ where $e~:~\Edge_{U(G)}(x,y)$ and the walk $w~:~\mathsf{W}_{U(G)}(y,z)$ is of length
    $n$. The following cases need to be considered concerning with the
    equality $x = y$.
\begin{enumerate}
    \item If $x = y$ then by the induction hypothesis applied to $w$,
    one obtains the normal form $\mathsf{nf}(w)$ of the walk $w$,
    along with $r : w \rightsquigarrow~\mathsf{nf}(w)$ and $h_1 : \| w
    \sim_{\mathcal{M}} \mathsf{nf}(w)\| $. We ask if $x = z$ to see if
    $p$ is a loop.
    \begin{enumerate}
        \item \label{item-x-is-z} If $x = z$ then the walk $p$ reduces
        to the trivial walk $\langle x \rangle$ by $\xi_1$. By
        applying $\mathcal{M}$ to the quasi-simple walk
        $\mathsf{nf}(w)$ and $\langle x \rangle$, $h_2~:~\|~\mathsf{nf}(w)~\sim_{\mathcal{M}}~\langle~z~\rangle~\|$ is
        obtained. It remains to show that $p$ is homotopic to $\langle
        x \rangle$. Because being homotopic is a proposition, the
        propositional truncation in $h_1$ and $h_2$ can be eliminated
        to get access to the corresponding homotopies. The required
        walk homotopy is as follows.
        \begin{equation*}
        \begin{array}{ll}
        p \equiv (e ⊙ w) &\\
        {\color{white} p}\sim_{\mathcal{M}} e ⊙ \mathsf{nf}(w)  &(\mbox{By \Cref{lem:whiskering} and }h_1)\\
        {\color{white} p}\sim_{\mathcal{M}} e ⊙ \langle z \rangle &(\mbox{By \Cref{lem:whiskering} and }h_2)\\
        {\color{white} p}\sim_{\mathcal{M}} \langle x \rangle &(\mbox{By \Cref{lem:loop-edges-homotopic-to-trivial-walks} applied to }\mathcal{M}).\\
        \end{array}
        \end{equation*}

        \item If $x \neq z$ then the walk $p$ reduces to
        $\mathsf{nf}(w)$ by the following calculation using
        $h_1$.
        \begin{equation*}
            \begin{array}{ll}
            p \equiv (e ⊙ w) &\\
            {\color{white} p}\equiv (e ⊙ ⟨ x ⟩) \cdot w  &(\mbox{By def. of walk composition})\\
            {\color{white} p}\rightsquigarrow^{*} w  &(\mbox{By }\xi_{3})\\
            {\color{white} p}\rightsquigarrow^{*} \mathsf{nf}(w) &(\mbox{By }r\mbox{}).\\
            \end{array}
        \end{equation*}
    \end{enumerate}
    \item
    \label[proof]{proof:split-at-in:thm:hom-normalisation} If $x \neq
    y$, then we split $w$ at $x$ using \Cref{def:view-splitat}. Hence,
    two cases have to be considered: whether $x$ is in $w$ or not, see
    \Cref{def:type-splitat}.
    \begin{enumerate}
        \item If $x \in w$, then, for every node $k$ in $G$, there are walks
        $w_1~:~\mathsf{W}_{U(G)}(y,k)$ and
        $w_2 : \mathsf{W}_{U(G)}(k,z)$ such that
        $\gamma : w = w_1 \cdot w_2$, along with evidence that $x \not
        \in w_1$ by \Cref{def:view-splitat}. By the induction
        hypothesis applied to $w_1$ and to $w_2$, we obtain the normal
        forms $\mathsf{nf}(w_1)$ and $\mathsf{nf}(w_2)$, and the terms 
        $r_i : w_i \rightsquigarrow \mathsf{nf}(w_i)$ and
        $h_i : \| w_i \sim_{\mathcal{M}} \mathsf{nf}(w_i) \|$ for
        $i=1,2$. The following cases concern with whether $x = z$ or
        not.
        \begin{enumerate}
            \item If $x = z$, the walk $p$ reduces to $\langle x
            \rangle$ by the rule $\xi_1$. To show that $p$ is
            homotopic to $\langle x \rangle$, let $s_1$ and $s_2$ of
            type, respectively, $\| p \sim_{\mathcal{M}} \mathsf{nf}(w_2) \|$ and
            $\|~\mathsf{nf}(w_2) \sim_{\mathcal{M}} ⟨ x ⟩ \|$, as given below.
            Assuming one has the terms $s_1$ and $s_2$, by elimination of
            the propositional truncation and the transitivity property
            of walk homotopy with $s_1$ and $s_2$, the required
            conclusion follows. The walk homotopy $s_1$ is as follows.

            \begin{equation*}\label{eq:suble-case}
                \begin{array}{ll}
                p \equiv \ (e ⊙ w) &\\
                {\color{white} p} \sim_{\mathcal{M}} e ⊙ (w_1 \cdot w_2)  &(\mbox{By the equality }\gamma)  \\
                {\color{white} p} \sim_{\mathcal{M}} (e ⊙ w_1) \cdot w_2  &(\mbox{By assoc. property of }(\cdot))\\
                {\color{white} p} \sim_{\mathcal{M}} (e ⊙ \mathsf{nf}(w_1)) \cdot \mathsf{nf}(w_2) &(\mbox{By \Cref{lem:whiskering}, } h_1\mbox{, and }h_2)\\
                {\color{white} p} \sim_{\mathcal{M}} \langle x \rangle \cdot \mathsf{nf}(w_2) &(\mbox{By the homotopy from }h_4)\\
                {\color{white} p} \sim_{\mathcal{M}} \mathsf{nf}(w_2) &(\mbox{By definition}),\\
                \end{array}
            \end{equation*}
            where $h_4 : \| (e ⊙ \mathsf{nf}(w_1)) \sim_{\mathcal{M}} \langle x \rangle\|$
            is given by applying the map $\mathcal{M}$ to
            the quasi-simple walks, $(e~⊙~\mathsf{nf}(w_1))$ and
            $\langle x \rangle$. The walk $(e~⊙~\mathsf{nf}(w_1))$ is
            quasi-simple by \Cref{lem:e-simple-is-simple}. Also, note that
            $x~\not \in~\mathsf{nf}(w_1)$ by \Cref{lem:nf} and the
            assumption $x~\not \in w_1$. Finally, the remaining walk homotopy
            $s_2$ is obtained by applying $\mathcal{M}$ to the quasi-simple
            walks, $\mathsf{nf}(w_2)$ and the trivial walk at $x$.

            \item If $x \neq z$, then the walk $p$ reduces to
            $\mathsf{nf}(w_2)$ by the reduction reasoning in
            \Cref{eq:p-reduces-w2}. As the walk $\mathsf{nf}(w_2)$ is
            in normal form, it remains to show that $p$ is homotopic
            to $\mathsf{nf}(w_2)$. However, the reasoning is similarly
            to \Cref{eq:suble-case}.

            \begin{equation}\label{eq:p-reduces-w2}
            \begin{array}{ll}
                p \equiv \ (e ⊙ w) &\\
                {\color{white} p} \rightsquigarrow^{*} e ⊙ (w_1 \cdot w_2)  &(\mbox{By splitting }w\mbox{ using \Cref{def:view-splitat}}) \\
                {\color{white} p} \rightsquigarrow^{*} (e ⊙ w_1) \cdot w_2  &(\mbox{By assoc. property of }(\cdot))\\
                {\color{white} p} \rightsquigarrow^{*} \langle x \rangle \cdot w_2  &(\mbox{By }\xi_{2}\mbox{ applied to the loop }(e ⊙ w_1))\\
                {\color{white} p} \rightsquigarrow^{*} w_2  &(\mbox{By definition of walk composition})\\
                {\color{white} p} \rightsquigarrow^{*} \mathsf{nf}(w_2) &(\mbox{By the induction hypothesis}).\\
            \end{array}
            \end{equation}
            
        \end{enumerate}

        \item Otherwise, there is evidence that $x~\not \in w$. By the
        induction hypothesis applied to $w$, the walk $\mathsf{nf}(w)$
        is obtained, along with a reduction
        $r~:~w~\rightsquigarrow~\mathsf{nf}(w)$ and evidence
        $h~:~\|~w~\sim_{\mathcal{M}}~\mathsf{nf}(w)~\|$. The proof is
        by structural induction on the walk $\mathsf{nf}(w)$.
        \begin{enumerate}
        \item If $\mathsf{nf}(w)$ is the trivial walk
        $\langle y \rangle$, then the walk $p$ reduces either to
        $\langle x \rangle$, if $x~=~z$, or to the walk
        $(e ⊙ \langle z \rangle)$, if $x \neq z$. Either way, it is
        possible to construct the corresponding homotopies, similarly
        as for \Cref{item-x-is-z}.

        \item If the walk $\mathsf{nf}(w)$ is the composite walk $(u~⊙~v)$ for
        $u : \Edge_{U(G)}(y, y')$, $v : \mathsf{W}_{U(G)}(y',z)$ and nodes
        $y',z~:~\Node_{G}$, then we ask if $x~=~z$.
\begin{itemize}
    \item If $x=z$ then the walk $p$ reduces to the trivial walk
    $\langle x \rangle$ by $\xi_1$. It remains to show that the walk
    $(e~⊙~\mathsf{nf}(w))$ is homotopic to $\langle x \rangle$. To see
    this, the spherical property of the map $\mathcal{M}$ is applied.
    Note that the walk $(e~⊙~\mathsf{nf}(w))$ is quasi-simple by
    \Cref{lem:e-simple-is-simple}, as $x~\not \in~\mathsf{nf}(w)$ by
    \Cref{lem:nf} applied to the assumption $x \not \in w$.

    \item If $x\neq z$ then the walk $p$ reduces to the walk
    $(e~⊙~\mathsf{nf}(w))$ by $\xi_2$. By the propositional truncation
    elimination applied to the evidence of \Cref{lem:whiskering} and
    to the homotopy $h$, one can obtain evidence that the walk
    $(e ⊙ w)$ is homotopic to $(e ⊙ \mathsf{nf}(w))$. It remains to show that the composite walk
    $(e ⊙ \mathsf{nf}(w))$ is in normal form. By \Cref{lem:e-simple-is-simple},
    this walk is quasi-simple. By case
    analysis on the possible reductions using \Cref{def:loop-reduction-relation}, one
    proves that this walk does not reduce. Therefore, $(e ⊙ \mathsf{nf}(w))$ is in normal form.
    {\qedhere}
\end{itemize}
        \end{enumerate}
    \end{enumerate}
    \end{enumerate}
\end{itemize}
\end{linkproof}

\begin{corollary}\label{lem:two-spherical-map-definition-are-equivalent}
    The two spherical map definitions, \Cref{def:spherical-map} and
    \Cref{def:spherical-map-simple}, are equivalent when considering
    graphs with discrete set of nodes.
\end{corollary}

\begin{linkproof}[]{https://jonaprieto.github.io/synthetic-graph-theory/CPP2022-paper.html\#2999}
The definitions in question are propositions. Thus, it is only
necessary to show that they are logically equivalent.
\begin{enumerate}
\item Every spherical map by \Cref{def:spherical-map-simple} is a
spherical map with additional data in the sense of
\Cref{def:spherical-map}
\item Let $\mathcal{M}$ be a spherical map by
\Cref{def:spherical-map-simple}. To see $\mathcal{M}$ also satisfies
\Cref{def:spherical-map}, let $w_1$ and $w_2$ be two quasi-simple
walks from $x$ to $y$. We must now exhibit evidence that $w_1$ is
homotopic to $w_2$. By \Cref{thm:hom-normalisation}, a walk homotopy
$h_1$ between $w_1$ and the normal form $\mathsf{nf}(w_1)$ exists.
Similarly, one can obtain a term $h_2$ of type $\| w_2 \sim_{\mathcal{M}} \mathsf{nf}(w_2) \|$. 

\begin{equation}\label{eq:eq-spherical-notions}
\begin{array}{ll}
w_{1}                   \sim_{\mathcal{M}} \mathsf{nf}(w_{1})  &(\mbox{By } h_1\mbox{ from \Cref{thm:hom-normalisation}}) \\
{\color{white} w_{1} }  \sim_{\mathcal{M}} \mathsf{nf}(w_{2})  &(\mbox{By } h_3\mbox{ from \Cref{def:spherical-map-simple}})\\
{\color{white} w_{1} }  \sim_{\mathcal{M}} w_{2}               &(\mbox{By } h_2\mbox{ from \Cref{thm:hom-normalisation}}).\\
\end{array}
\end{equation}

On the other hand, recall that walks in normal form are
quasi-simple walks by definition. Therefore, it is possible to get
$h_3 : \| \mathsf{nf}(w_1) \sim_{\mathcal{M}}\mathsf{nf}(w_2) \|$ by
applying the spherical property of the map $\mathcal{M}$ to
$\mathsf{nf}(w_1)$ and $\mathsf{nf}(w_2)$. By the elimination of the
propositional truncation applied to $h_1$, $h_2$, and $h_3$, the
required evidence of a homotopy between $w_1$ and $w_2$ can be
obtained, as stated in \Cref{eq:eq-spherical-notions}.\qedhere

\end{enumerate}
\end{linkproof}

\section{Related Work}\label{sec:related-work}

In other areas of mathematics unrelated to type theory, considering
homotopy for graph-theoretical concepts, for example, is not new.
There are several proposals of the concept of homotopy for graphs
using a few discrete categorical constructions \citep{Grigoryan2014}.
Many of these constructions use the \(\times\mbox{-}\text{homotopy}\)
notion, defined as a relation based on the categorical product of
graphs in the Cartesian closed category of undirected graphs. Since a
walk of length \(n\) in a graph \(G\) is simply a morphism between a
path graph \(P_n\) into \(G\), the notion of homotopy for walks is
there defined as homotopy between graph homomorphisms. The looped path
graph \(I_n\) is used to define the homotopy of these morphisms---in a
manner similar to the interval \([0,1]\) for the concept of homotopy
between functions in homotopy theory. As a source of more results, it
is possible to endow the category of undirected graphs with a
\(2\)-category structure by considering homotopies of walks as
\(2\)-cells, as described by Chih and Scull \citep{Chih2020}.

On the reduction relation on walks and spherical maps, this work is
related to polygraphs used in the context of higher-dimensional
rewriting systems. Recent works by Kraus and von Raumer
\citep{Kraus2021Rewriting, Kraus2020} use ideas in graph theory,
higher categories, and abstract rewriting systems to approximate a
series of open problems in HoTT. In the same vein, the internalisation
of rewriting systems and the implementation of polygraphs in Coq by
Lucas \citep{lucas2020abstract, lucas:hal-02385110} was found to be
related to Kraus and von Raumer's approach. One fundamental object in
the work by the authors mentioned above is that of an \(n\)-polygraph,
also called \emph{computad}.

An \(n\)-polygraph is a (higher dimensional) structure that can serve,
for example, to analyse reducing terms to normal forms and comparing
reduction sequences on abstract term rewriting systems. The following
is a possible correspondence to relate these ideas within the context
of our work. The notion of a \(1\)-polygraph
\citep[§2]{Kraus2021Rewriting}---which is given by two sets
\(\Sigma_0\) and \(\Sigma_1\), and two functions
\(s_0, t_0 : \Sigma_1 \to \Sigma_0\)---is
\href{https://jonaprieto.github.io/synthetic-graph-theory/lib.graph-definitions.Alternative-definition-is-equiv.html}{equivalent}
to the type of graphs in \Cref{def:graph}. An \emph{object} is a node,
a \emph{reduction step} is an edge, and a \emph{reduction sequence}
\(a \rightsquigarrow^* b\) is a walk between nodes \(a\) to \(b\). A
(closed) \emph{zig-zag} is a (cycle) walk in the symmetrisation of the
graph representing the reduction relation. A (generalised)
\(2\)-polygraph \citep[Def. 25]{Kraus2021Rewriting} consists of a type
\(A\), a set of reduction steps on \(A\), and all rewriting steps
between zig-zags. Then, the notion of \(2\)-polygraph on \(A\) will
correspond to a graph \(G\) representing the type \(A\) with the set
of all walks in \(G\) and the collection of walk homotopies in the
symmetrisation \(U(G)\) for a given combinatorial map.

Using the previous interpretation for polygraphs, one may state that a
graph with a spherical map holds properties such as
\emph{terminating}, \emph{closed under congruence}, \emph{cancels
inverses}, and it has a \emph{Winkler-Buchberger structure} \citep[Eq.
32-35]{Kraus2021Rewriting}. The related concept of
\emph{homotopy basis} of a \(2\)-polygraph \citep[Def.
28]{Kraus2021Rewriting} may be seen as the set obtained from
\Cref{def:spherical-map} without using propositional truncation in the
corresponding type.

On the other hand, \emph{Noetherian induction for closed zig-zags}
\citep[§ 3.5]{Kraus2021Rewriting} addresses a similar issue we
investigated herein. In this work, we found out that to prove certain
properties, as the normalisation theorem in
\Cref{thm:hom-normalisation} for graphs with a spherical map and a
discrete set, it was only necessary to consider (cycle) walks without
inner loops. One can prove other properties related to walk homotopies
for graphs with spherical maps, not only considering the property on a
cycle walk but any walk. This approach relies on the machinery of
quasi-simple walks in \Cref{sec:quasi-simple} and the loop reduction
relation on walks in \Cref{sec:loop-reduction-relation}. Our
loop-reduction relation is likely \emph{locally confluent} \citep[§
3.3]{Kraus2021Rewriting}, but without uniqueness of normal forms. We
leave the proof of these properties as future work because they were
not required here. We will also investigate in-depth the extent to
which the constructions given by Kraus and von Raumer, as well as by
Lucas, are not only related but applicable to our main project of
graph theory in HoTT \citep{agdaformalisation}.

Finally, on the computer formalisation side, the use of formal systems
to formalise graph-theoretical results on the computer is not a
novelty. The proof of the four-colour theorem (FCT) in Coq by Gonthier
\citep{Gonthier2008} is one famous example that works with
hypermaps---a similar notion to combinatorial maps, as defined in
\Cref{def:graph-map}. However, both the type theory and the goal of
the constructions are substantially different from our exposition.
There are other relevant projects in the field and extensive libraries
of graph theory in Coq \citep{doczkal2020}, Isabelle/HOL
\citep{Noschinski2015}, and Lean \citep{halltheoremLean}. However, to
the best of our knowledge, few efforts use a proof-relevant dependent
type theory like HoTT and a proof assistant like Agda. We find only
the work mentioned earlier by Kraus and von Raumer
\citep{Kraus2021Rewriting, Kraus2020} to be related to our Agda
development; their work contains a
\href{https://gitlab.com/fplab/freealgstr}{formalisation} of their
results in a version of the proof assistant Lean compatible with HoTT.

In other formal developments like
\href{https://unimath.github.io/doc/UniMath/d4de26f//UniMath.Combinatorics.Graph.html}{the
HoTT Coq Library} \citep{HoTTCoq},
\href{https://unimath.github.io/doc/UniMath/d4de26f//UniMath.Combinatorics.Graph.html}{the
UniMath Library} \citep{UniMath}, and
\href{https://github.com/agda/cubical/tree/9bedca9b94d41b8efe9fb541ceb0d158464e9497/Cubical/Data/Graph}
{the Standard Library of Cubical Agda}, only the basic definitions are
available (e.g.~the type of graphs, graph homomorphisms, and
diagrams). Future work might involve porting our development into one
of these libraries.

\section{Concluding Remarks}\label{sec:conclusions}

This work proves some non-trivial results for directed multigraphs
using a proof-relevant approach in the language of homotopy type
theory. This work supports an ongoing project to define planarity
criteria and other concepts of graph theory in HoTT
\citep{types2019abstract, planarityHoTT} formalised in Agda
\citep{agdaformalisation}.

In our formalisation, each definition and theorem presented herein is
related to a term in the proof assistant Agda. This approach was
helpful to reveal and confirm that only a subset of HoTT was necessary
to perform all the proofs in this development. Precisely, we only need
the intensional Martin--Löf type theory equipped with universes,
function extensionality, and propositional truncation. No other higher
inductive type is required. It is worth noting that without
considering propositional truncation, it would not have been possible
to define our main theorems. The propositional truncation allows us to
model the mere existence of an object in theory correctly.

This work's primary contributions are \Cref{thm:hom-normalisation},
and especially \Cref{lem:two-spherical-map-definition-are-equivalent}.
In summation, \Cref{thm:hom-normalisation} states that we can
normalize any walk to a normal form that is walk-homotopic to it
whenever the graph has a discrete node set and is embedded in the
sphere. On the other hand,
\Cref{lem:two-spherical-map-definition-are-equivalent} establishes an
equivalence between two definitions of embeddings in the sphere for
graphs with a discrete node set. Except for this last result, the
machinery shown in this paper was utterly unexpected and developed
solely to find evidence for our initial conjecture. For characterising
embeddings of finite graphs in the sphere, one needs to consider only
the finite set of walks without internal loops. Using the results
given herein, one can devise a (brute-force) algorithm to determine
whether an embedding is spherical or not. Future work will be devoted
to implementing this algorithm. To the best of our knowledge, we
provided the minimum to demonstrate
\Cref{thm:hom-normalisation,thm:normalisation,lem:two-spherical-map-definition-are-equivalent}.

\begin{acks}
The author thanks Håkon R. Gylterud for very helpful discussions on
various issues related to this paper. Thanks to Marc Bezem and the
anonymous reviewers for the comments, references and suggestions that
improved this document. Thanks to the Department of Informatics at the
University of Bergen for funding this research. Last but not least,
thanks to the Agda developer team for providing and maintaining the
proof assistant used to check the results of this work.
\end{acks}

\bibliography{ref.bib}
\end{document}